\documentclass[twocolumn,superscriptaddress,floatfix,preprintnumbers, nofootinbib,hyperref]{revtex4} %
\usepackage{amsmath, amsfonts, amssymb,color}
\usepackage[normalem]{ulem}

\usepackage{graphicx}%  Default Latex 2eps package for embedding figures
\usepackage{rotating}
\usepackage{epsfig}
\usepackage{bm}
\usepackage[usenames,dvipsnames]{xcolor}
\usepackage{color}

\def\he4{$^4$He}
\def\h2{$^2$H}

\DeclareMathOperator{\cm}{cm}
\DeclareMathOperator{\GeV}{GeV}
\DeclareMathOperator{\eV}{eV}

\DeclareMathOperator{\keV}{keV}

\DeclareMathOperator{\diag}{diag}

\makeatletter
\newcommand{\dalembertian}{\mathop{\mathpalette\dalembertian@\relax}}
\newcommand{\dalembertian@}[2]{%
  \begingroup
  \sbox\z@{$\m@th#1\square$}%
  \dimen0=\fontdimen8
    \ifx#1\displaystyle\textfont\else
    \ifx#1\textstyle\textfont\else
    \ifx#1\scriptstyle\scriptfont\else
    \scriptscriptfont\fi\fi\fi3
  \makebox[\wd\z@]{%
    \hbox to \ht\z@{%
      \vrule width \dimen0
      \kern-\dimen0
      \vbox to \ht\z@{
        \hrule height \dimen0 width \ht\z@
        \vss
        \hrule height 2\dimen0
      }%
      \kern-2.5\dimen0
      \vrule width 2.5\dimen0
    }%
  }%
  \endgroup
}
\makeatother

\begin{document}

\title{Production of axion-like particles from photon conversions in large-scale solar magnetic fields
}

\author{Ersilia Guarini}
\affiliation{Dipartimento Interateneo di Fisica ``Michelangelo Merlin'', Via Amendola 173, 70126 Bari, Italy}

\author{Pierluca Carenza}
\affiliation{Dipartimento Interateneo di Fisica ``Michelangelo Merlin'', Via Amendola 173, 70126 Bari, Italy}
\affiliation{Istituto Nazionale di Fisica Nucleare - Sezione di Bari, Via Orabona 4, 70126 Bari, Italy}

\author{Javier Gal\'an}
\affiliation{Grupo de Fisica Nuclear y Astroparticulas, Departamento de Fisica Teorica,
Universidad de Zaragoza C/ P. Cerbuna 12 50009, Zaragoza, Spain }

\author{Maurizio~Giannotti}
\affiliation{Physical Sciences, Barry University, 11300 NE 2nd Ave., Miami Shores, FL 33161, USA}

\author{Alessandro~Mirizzi} 
\affiliation{Dipartimento Interateneo di Fisica ``Michelangelo Merlin'', Via Amendola 173, 70126 Bari, Italy}
\affiliation{Istituto Nazionale di Fisica Nucleare - Sezione di Bari, Via Orabona 4, 70126 Bari, Italy}

\date{\today}

\begin{abstract}
The Sun is a well-studied astrophysical source of axion-like particles (ALPs), 
  produced mainly through the Primakoff process.
Moreover,  in the Sun there exist large-scale magnetic fields that catalyze an additional 
ALP production via a coherent conversion of thermal photons. 
We study this contribution to the solar ALP emissivity, typically neglected in 
previous investigations. 
%\xout{In order to characterize this flux, we  solve the kinetic equations of the photon-ALP system in large-scale solar $B$-fields}. 
Furthermore, we discuss additional bounds on the ALP-photon coupling
from energy-loss arguments, and the detection perspectives of this new ALP flux at future helioscope and dark matter experiments.
\end{abstract}

\maketitle

\section{Introduction}

Axion-like particles (ALPs) are ultralight pseudo-scalar bosons $a$ with a two-photon vertex $a \gamma \gamma$,
predicted by several extensions of the Standard Model (see~\cite{Jaeckel:2010ni,DiLuzio:2020wdo} for comprehensive reviews).
  The two-photon coupling allows the conversion of ALPs into photons, $a\leftrightarrow\gamma$, in external electric or magnetic fields.
In stars, this leads to  the Primakoff process that induces the production of low mass
ALPs in the microscopic electric fields of nuclei and electrons. 
An ALP flux would then cause a novel source of energy-loss in stars, altering their evolution.
In this context, the strongest bound comes from helium-burning stars in globular clusters, giving $g_{a\gamma} <  6.6 \times 10^{-11}$~GeV${}^{-1}$ for 
$m_a \lesssim 1$~keV~\cite{Ayala:2014pea}. 
Another interesting possibility is the conversion in a macroscopic field, usually a large scale magnetic field.
In this case,
the momentum transfer is small, the interaction is coherent over a large
distance, and the conversion is best viewed as an axion-photon oscillation
phenomenon in analogy to neutrino flavor oscillations. 
This  effect is exploited to search for generic ALPs in light-shining-through-the-wall experiments (see e.g.~the ALPS~\cite{Ehret:2010mh} and
 OSQAR~\cite{Ballou:2015cka} experiments), for solar ALPs (see e.g.~the CAST experiment~\cite{Arik:2011rx,Arik:2013nya}) and for ALP dark 
 matter~\cite{Arias:2012az} in micro-wave cavity experiments (e.g.,~the ADMX experiment~\cite{Braine:2019fqb}). 
 In particular, the solar ALP search in a helioscope, like CAST,  exploits the production of an ALP flux with $E\sim {\mathcal O}$(few) keV via Primakoff process in the Sun core and its back-conversion into $X$--rays in the large-scale magnetic field of the detector.
 The absence of an ALP signal allows one to get the best experimental bound on the photon-ALP coupling, $g_{\rm a\gamma}\lesssim 6.6 \times 10^{-11}$~GeV$^{-1}$  for 
$m_a\lesssim 0.02$~eV~\cite{Anastassopoulos:2017ftl}, comparable with the one placed from helium-burning stars.

The Sun can be a source of intense magnetic fields that can be relevant for ALP conversions. 
Notably, in~\cite{Carlson:1995xf} it was studied the possibility
to trigger ALP conversions in $X$--rays in the intense  magnetic fields of the sunspots on the Sun surface.
Observations of 
the Soft X--rays Telescope (SXT) on the Yohkoh satellites allows to obtain bounds on 
$g_{\rm a\gamma}\lesssim  {\mathcal O} (10^{-10})$~GeV$^{-1}$. 
Presumably,
this bound can be strengthened with a dedicated Sun observation by the current NuStar satellite experiment~\cite{vogel}.
Furthermore, even though in the Standard Solar Models (SSMs)~\cite{Bahcall:2004pz,Serenelli:2009yc}  the Sun is assumed as a quasi-static environment, seismic solar models
have been developed including large-scale magnetic fields in different regions of the solar interior~\cite{TurckChieze:2001ye,Couvidat:2003ba}.
The presence of these $B$-fields may trigger conversions of the thermal photons into ALPs, creating an additional ALP flux besides
the one produced by the Primakoff conversions. 
Some preliminary characterization of this flux has been presented in talks~\cite{talkmirizzi,talkgalan}. 
However, a detailed calculation
is still lacking in the literature. 
Only recently there appear dedicated studies of the ALP flux produced in the solar interior via
conversions in the $B$-fields of longitudinal plasmons~\cite{Caputo:2020quz,OHare:2020wum} (see also~\cite{Mikheev:1998bg,Tercas:2018gxv,Mendonca:2019eke}).
Our work complements these results by taking into account the  conversions into ALPs  for transverse, as well as longitudinal photon modes in the solar plasma.

The plan of our paper is as follows. 
In Sec.~\ref{sec:Solar_B} we describe the model of the solar $B$-fields we will use to characterize the photon 
conversions into ALPs. 
In Sec.~\ref{sec:Photon_ALP_conversions} we revise the conversions of photon into ALPs for longitudinal and transverse modes.
In Sec.~\ref{sec:ALP_rates}  we solve the ALP-photon kinetic equations in order to compute the ALP production
rate, taking into account the photon absorption in the solar plasma.
In Sec.~\ref{sec:ALP_fluxes} we calculate the solar ALP fluxes from magnetic conversions.
In Sec.~\ref{sec:Energy-loss} we present a new bound on the ALP-photon coupling $g_{a\gamma}$ 
from energy-loss argument associated with 
ALPs emitted from photon conversions in $B$-fields. 
In Sec.~\ref{sec:detection}
we discuss the detection perspectives for these fluxes. 
Finally, in  Sec.~\ref{sec:conclusions} we summarize our results and we conclude.
There follow Appendix A, where we give details of the calculation of conversion probabilities into ALPs for transverse and longitudinal photons;
Appendix B, where we describe the solution of the ALP-photon kinetic equations;
and Appendix C, where we present the thermal field theory approach. We show that 
the kinetic and the thermal field theory approach lead to the same ALP production rates.

%%%%%%%%%%%%%
\section{Solar magnetic fields}
\label{sec:Solar_B}
%%%%%%%%%%%%%%

The magnetic field of the Sun is most important in three different regions, 
the \textit{radiative} zone ($r \lesssim 0.7 \; R_\odot$), the exterior (\textit{convective}) zone ($r \gtrsim 0.9 \; R_\odot$), 
and the intermediate region between these two, called the \textit{tachocline} ($ r \sim 0.7 \; R_\odot$).
{Here, we are using the standard notation $ R_{\odot} =6.9598\times 10^{10}$~m for the solar radius~\cite{Serenelli:2009yc}.} 
In our work for simplicity we assume spherical symmetry for the solar magnetic field, described by a radial profile $B(r)$
(see the seismic model in~\cite{Dzitko:1995zz}  for a  toroidal magnetic field).

The radiative zone (for $r\leq r_0 =0.712 R_\odot$) is characterized by the following profile~\cite{gough}
\begin{equation}
B(r)=  K_\lambda \left(\frac{r}{r_0} \right)^2 \left[1- \left( \frac{r}{r_0} \right)^2 \right]^\lambda \,B_{\rm{rad}}\,\ ,
\label{eq:radiative}
\end{equation} 
%where $B_{\rm{rad}}$ is the amplitude of the field
where $K_\lambda=(1 + \lambda)(1+ 1/\lambda)^\lambda $, with $\lambda=1 + 10 \, r_0/R_\odot$, and
  $1 \times 10^7 \; \rm{G} \lesssim B_{\rm {rad}} \lesssim 3 \times 10^7 \; \rm{G}$. 
This range was determined by Couvidat \textit{et al.} \cite{Couvidat:2003ba}. 
They used the precision on solar sound speed and density to rule out fields with intensity $B_0 \sim 10^4 \; \rm{T}$ 
and arguments on the solar oblateness to set the upper value. 

\begin{figure}[t]
\centering
\includegraphics[width=\columnwidth]{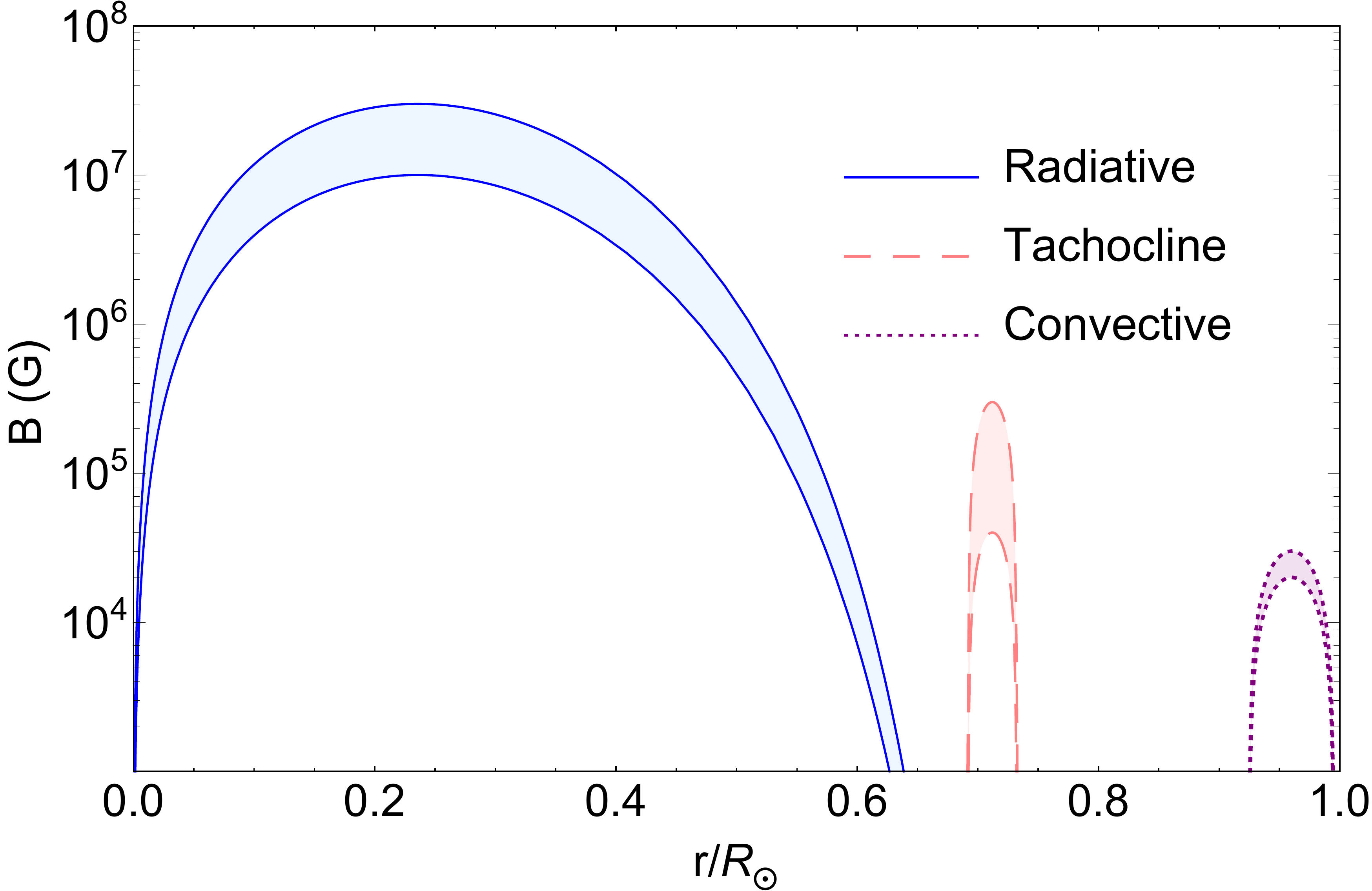}
\caption{Profile of solar magnetic field $B(r)$ as a function of 
{the normalized}
	radius $r/R_\odot$. In each region we take the following ranges: $B_{\rm{rad}} \in [1,3] \times 10^7 \; \rm{G}$ \cite{Couvidat:2003ba}, $B_{\rm{tach}} \in [3, 5] \times 10^5\; \rm{G}$ \cite{Antia:2000pu} and $B_{\rm{tach}} \in [2, 3] \times 10^4 \; \rm{G}$ \cite{Antia:2000pu}.}
\label{fig:bfields}
\end{figure}

The field profile in the tachocline is  simulated as
%%%%
\begin{equation}
B(r)=
B_{\rm{m}} \left[1- \left( \frac{r-r_0}{d}\right)^2 \right] \,\ , \,\  \textrm{for} \,\ |r-r_0| \leq d \\
\label{eq:tachocline}
\end{equation}
%%%%
where $r_0=0.712 R_\odot$ is the center of the zone and $d$ is its half-width. 
As benchmark parameters in the tachocline we set  $d=0.02 \; R_\odot$, while $ 3 \times 10^5 \; \rm{G} \lesssim B_{\rm{m}} \equiv B_{\rm{tach}}\lesssim 5 \times 10^5 \; \rm{G}$. 
These bounds were set by Antia \textit{et al.} by the observation of the splittings of solar oscillation frequencies~\cite{Antia:2000pu}. 
Similarly, the field profile in the upper layers is simulated as in Eq.~(\ref{eq:tachocline}),
with $r_0=0.96 \; R_\odot$, $d=0.035 \; R_\odot$ and $ 2 \times 10^4 \; \rm{G} \lesssim B_{\rm{m}} \equiv B_{\rm{conv}}\lesssim  3 \times 10^4 \; \rm{G}$. 
These bounds were {also set in Ref.~\cite{Antia:2000pu},} from an analysis of the Global Oscillation Network Group (GONG).
%To summarize, we simulate magnetic fields with strength in the different regions given by~\cite{Antia:2000pu}
%%%%
%\begin{eqnarray}
% B_{\rm{rad}} &\in& [1;3] \times 10^7 \; \rm{G}\;\ ,  \nonumber \\
% B_{\rm{tach}} &\in& [3;5] \times 10^5\; \rm{G} \;\ ,  \nonumber \\
% B_{\rm{conv}} &\in& [2;3] \times 10^4 \; \rm{G} \;\ . 
 %\label{eq:lim} 
%\end{eqnarray}
%%%%
The radial profile of the solar magnetic field described above is shown in  Fig.~\ref{fig:bfields}.

%%%%%%%%%%%%%%%%%%
%%%%%%%%%%%%%%%%

%%%%%%%%%%%%%%%%%%%%%%
\section{Photon-ALP conversions in the Sun}
\label{sec:Photon_ALP_conversions}
%%%%%%%%%%%%%%%%%%%%%%

%..........
\subsection{Photon dispersion in a plasma}
\label{sec:dispersion}
%...............

The dispersion relation of a photon in a plasma has the form~\cite{Jackson}
%.........
\begin{equation}
\omega^2 - k^2 = \pi_{T,L}(\omega, k)\; ,
\end{equation}
%................
where $\omega$ and $k$ are the photon frequency and wave-number and 
$ \pi_{T,L}(\omega, k)$ are the projection of the photon polarization tensor for
the transverse (T) and longitudinal (L) modes, respectively.
In particular, for the energies we are interested in, 
the dispersion relation for  transverse photons (TP) becomes~\cite{Jackson}
%%%%
\begin{equation}
\omega^2 - k^2 \approx \omega_{\rm p}^2 \; ,
\end{equation}
%%%%
where
 %%%%
\begin{equation}
\omega_{\rm p}=\sqrt{ \frac{4 \pi \alpha n_e}{m_e}}
= 1.31 \times 10^{18} \sqrt{\frac{n_e}{10^{26} \; \textrm{cm}^{-3}}} R_\odot^{-1} \,\ ,
\label{eq:plasma}
\end{equation}
%%%%
is the plasma frequency, with $n_e$ the electron density,
the numerical expression referring to typical solar conditions, in units of the solar radius $R_{\odot}$.

Instead, the longitudinal mode,  the so-called  longitudinal plasmon (LP),  has a  dispersion relation~\cite{Jackson}
 %%%%
\begin{equation}
\biggl( \frac{\omega}{\omega_{\rm p}} \biggr)^2 =1+ \frac{3 p}{mn_e} \biggl( \frac{k}{\omega_{\rm p}}\biggr)^2 \; ;
\end{equation}
%%%%
where $p$ is the equilibrium pressure. 
In the typical conditions of the solar plasma, the second term on the right hand side is negligible. 
Therefore, in the Sun the dispersion relation for LP reduces to 
%...........
\begin{equation}
\omega^2 \approx \omega^2_{\rm p} \,\ .
\end{equation}
%..........
We now discuss how TP and LP mix with ALPs in a plasma. The derivation of the equations of motion for such a system is given in Appendix A.

%..............
\subsection{Transverse photons}
%..................

The ALP-photon interaction Lagrangian is given by~\cite{Raffelt:1987im}
%%%%
\begin{equation}
\mathcal{L}= - \frac{1}{4} g_{a \gamma} a F_{\mu \nu} \tilde{F}^{\mu \nu} \; ;
\label{eq:coupledlagrangiana}
\end{equation} 
%%%%
where $a$ is the ALP field,
$g_{a \gamma}$ is the ALP-photon coupling,  $F_{\mu \nu}$ is the electromagnetic field tensor and $\tilde{F}_{\mu \nu}= 1/2 \epsilon_{\mu \nu \rho \sigma} F^{\rho \sigma}$ its dual.
Equation (\ref{eq:coupledlagrangiana}) is responsible for the \emph{mixing} among ALP and photons.

Transverse photons  mix with ALPs only through an external \emph{transverse} magnetic field ${\bf B}_{\rm ext} \equiv {\bf B}_{T}$.
We denote with $A_\bot$ and $A_\parallel$ the components of the vector potential ${\bf A}$ perpendicular and parallel to ${\bf B}_{T}$ respectively.
Assuming a uniform magnetic field we can reduce the general $3 \times 3$ mixing problem into a $2 \times 2$ system involving only $A_\parallel$ and $a$,
described by a  Schr${\rm{\ddot o}}$dinger like equation \cite{Raffelt:1987im,Anselm:1987vj}
%%%%
\begin{equation}
i \partial_z 
\left(\begin{matrix} A_\parallel  \\  a \end{matrix} \right)
=
H_T
\begin{pmatrix}
A_\parallel \\
a
\end{pmatrix} \; ;
\end{equation}
%%%%
where the Hamiltonian for the transverse modes reads (up to an overall phase diagonal term)
%%%%
\begin{equation}
H_{T}= 
\begin{pmatrix}
{\omega^2_p}/{2 \omega} & g_{a \gamma} B_T /2 \\
g_{a \gamma} B_T /2 &{m^2_a}/{2 \omega}
\end{pmatrix} \; .
\label{eq:htransa}
\end{equation}
%%%%
The TP-ALP conversion probability after traveling a distance $z$ in a uniform magnetic field $B_T$ is given by~\cite{Raffelt:1987im}
%%%%
\begin{equation}
P(\gamma_T \rightarrow a) = (\Delta_{a \gamma}^T z)^2 \frac{\sin^2(\Delta^T_{\rm {osc}} z/2)}{(\Delta^T_{{\rm osc}} z/2)^2} \; ;
\label{eq:probconva}
\end{equation}
%%%%
where we have introduced
%%%%
\begin{eqnarray}
 \Delta_{a \gamma}^T&=& g_{a \gamma} B_{T}/2 \; ; \nonumber \\
 \Delta_{\rm {osc}}^T&=&\sqrt{4 {\Delta_{a \gamma}^T}^2+(\Delta_{\rm{p}}-\Delta_a)^2}  \,\ .
 \label{eq:delta_T}
\end{eqnarray}
%%%%
In solar units, 
%%%%
\begin{eqnarray}
 \Delta_{ a \gamma}^T &=& \frac{g_{a \gamma} B_{T}}{2} \nonumber \\
 &\simeq& 1.2 \times 10^{-1} \left( \frac{g_{a \gamma}}{10^{-11} \; \textrm{GeV}^{-1}} \right) \left( \frac{B_T}{3 \times 10^5 \; \rm{G}} \right) R_\odot^{-1} \,\ , \nonumber\\
\Delta_{\rm p} &=& -\frac{\omega^2_{\rm p}}{2 \omega} \simeq -2.46 \times 10^{17} \biggl( \frac{\omega}{\textrm{keV}} \biggr)^{-1} \biggl( \frac{n_e}{10^{26} \rm{cm}^{-3}}\biggr) R_\odot^{-1} \,\ , \nonumber\\
 \Delta_a &=& -\frac{m^2_a}{2 \omega} \simeq -1.76 \times 10^{14} \biggl( \frac{m_a}{10 \; \textrm{eV}} \biggr)^2 \biggl( \frac{\omega}{\textrm{keV}} \biggr)^{-1} R_\odot^{-1} \; \,\ . \nonumber\\
 \label{eq:transvparam}
 \end{eqnarray}
%%%%

%%%%%%%%%
\begin{figure}[t]
\centering
\vspace{-0.5cm}
\includegraphics[width=\columnwidth]{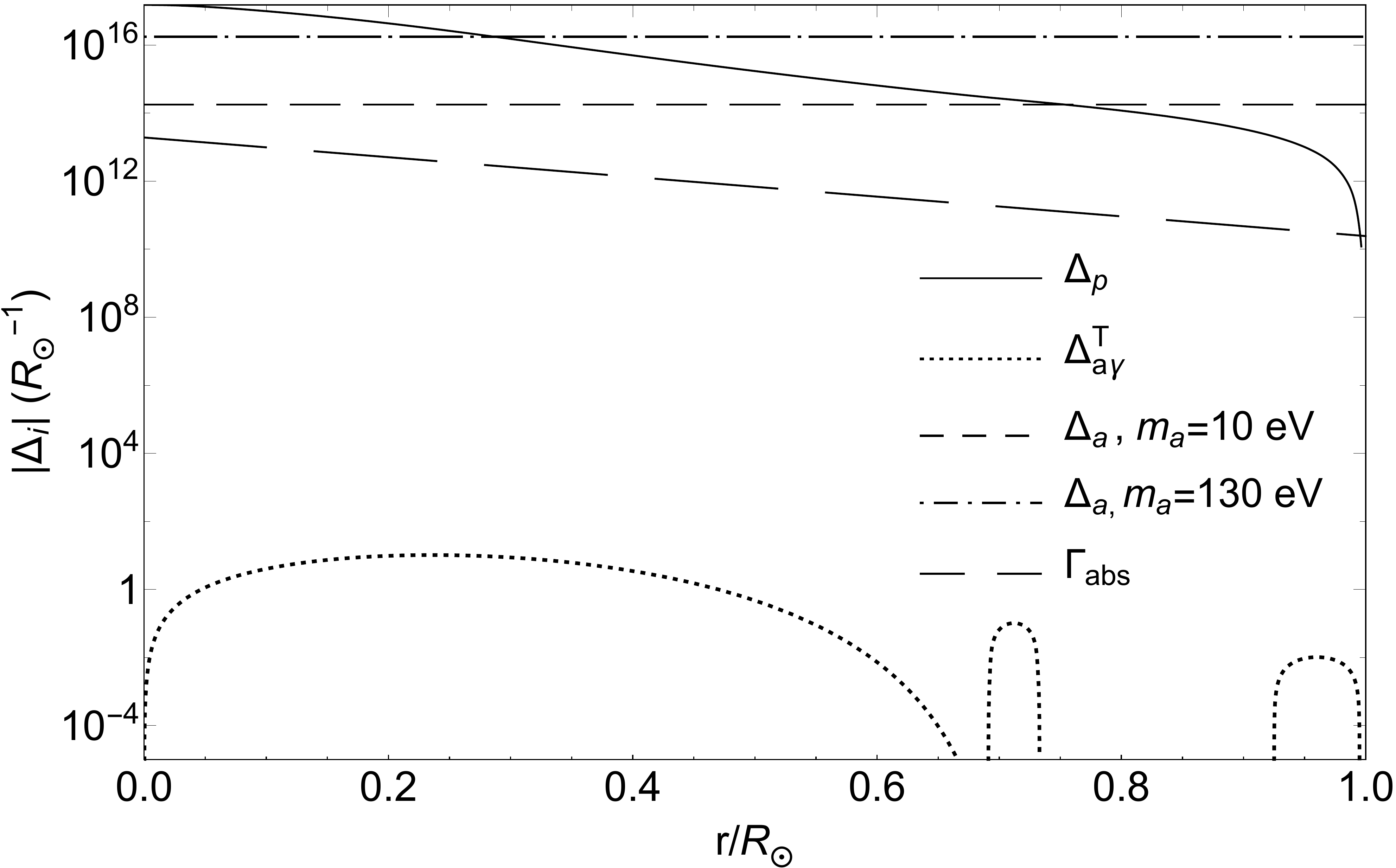}
\caption{Radial behavior of the  parameters for TP-ALP conversions [Eq.~(\ref{eq:transvparam})]  as a function of $r/R_\odot$.
}
\label{fig:oscpar}
\end{figure}
%%%%%%%

The radial behavior of these quantities in the Sun is shown in Fig.~\ref{fig:oscpar}. 
The plasma frequency has been characterized taking as reference the Solar Model AGSS09~\cite{Serenelli:2009yc}, which we will use 
as benchmark for our estimations. For the solar $B$-fields we used the model of 
 Sec.~\ref{sec:Solar_B}. 
 
\begin{figure}[t]
\centering
\includegraphics[width=\columnwidth]{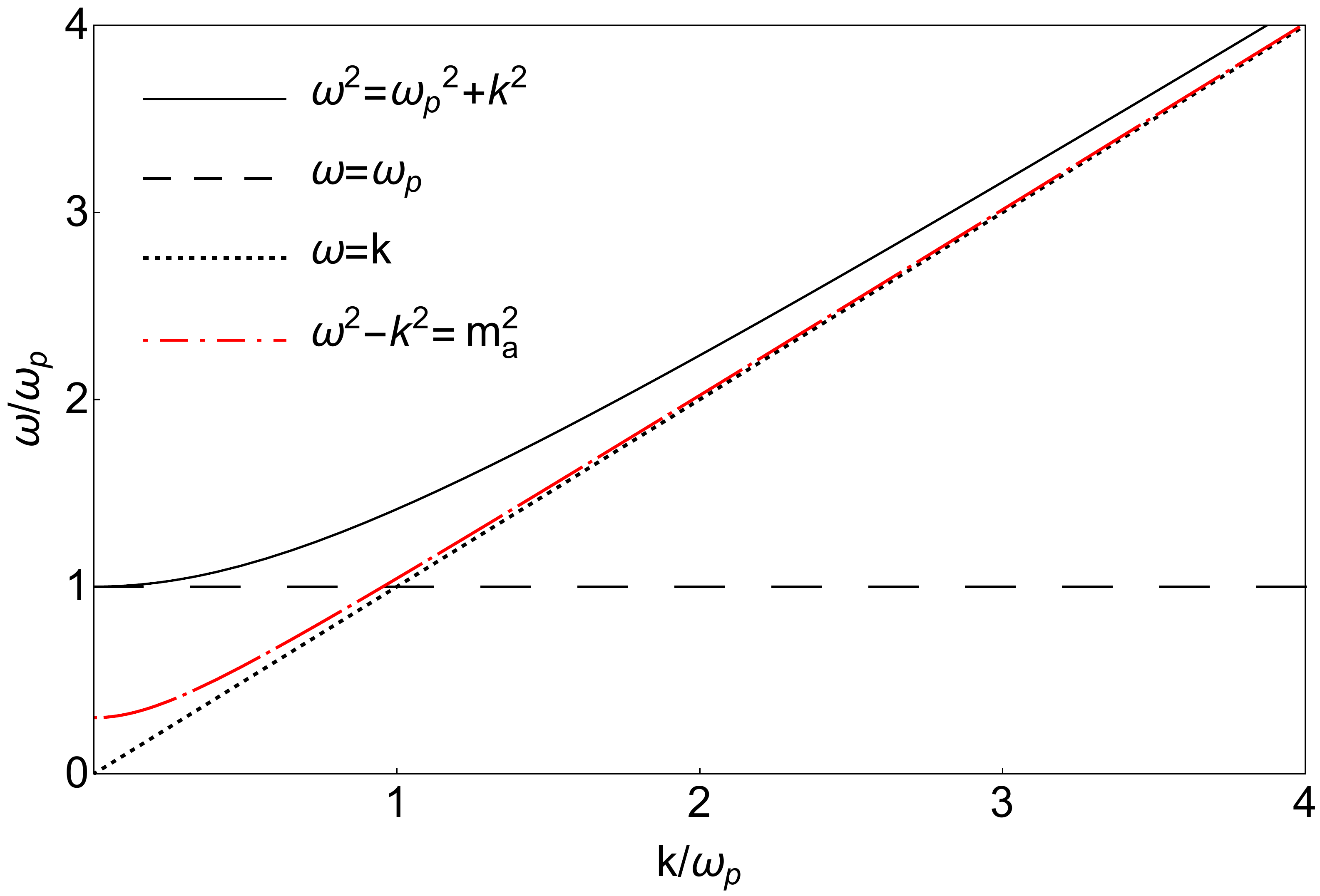}
\caption{Dispersion relation for an ALP with mass $m_a$ (dot-dashed curve). The ALP mass was assumed   large enough to  distinguish the dispersion relation  from that of an ordinary photon (dotted curve). The ALP dispersion crosses the one of the LP (dashed curve) for $\omega= \omega_{\rm {p}}$, where the resonance occurs. For TP, the only possibility of crossing between the ALP dispersion relation  and the transverse photon one (continuous curve) is $m^2_a=\omega^2_{\rm{p}}$.}
\label{fig:dispassione}
\end{figure}

 Remarkably, the TP-ALP conversion probability exhibits a \textit{resonant} behavior. 
Indeed, it is easy to see from Eq.~(\ref{eq:probconva})--(\ref{eq:delta_T})
that the probability is maximal when $\Delta_a=\Delta_{\rm {p}}$, i.e. when  $m^2_a=\omega^2_{\rm {p}}$. 
In this situation
 the ALP dispersion relation
(dot-dashed curve in Fig.~\ref{fig:dispassione}) matches the one of the TP (continuous curve).
 From Fig.~\ref{fig:oscpar}, one realizes that the resonance occurs in the radiative zone at $r \sim 0.3 R_\odot$ for $m_a \sim 100 \; \textrm{eV}$. Instead, the resonance in the tachocline occurs at $r \sim 0.7 R_\odot$ for $m_a \sim 10 \; \textrm{eV}$. 
 In the following we will take these values of ALP mass as benchmark for the calculation of resonant ALP production.
 In principle, we may have also a resonant conversion in the convective zone at  $r \sim 0.9 R_\odot$ for
 $m_a \sim 1 \; \textrm{eV}$. However, due to the lower local temperature and to the smaller magnetic field the resultant ALP flux would be smaller
 than the previous ones and we will neglect it hereafter.
 
%%%%%%%%%%%%
%%%%%%%%%%%%%%%%%%%%%%%%

Finally, we need to take into account that thermal photons are continuously emitted and re-absorbed in the Sun. 
This process is characterized by the transverse photon absorption coefficient rate $\Gamma_{\rm {abs}}$,  
defined as the inverse of the photon mean free path $\lambda_{\rm mfp}$.
An explicit calculation of this rate is presented in~\cite{Redondo:2013lna}.
 We take its numerical values, shown in Fig.~\ref{fig:oscpar},  
from~\cite{Krief:2016znd}.
One realizes that the photon absorption coefficient $\Gamma_{\rm {abs}}$ in the solar plasma is not negligible with respect to the other oscillation parameters.
Rather, it is always larger than  $\Delta_{a \gamma}^T$. 
Therefore, it needs to be included in the treatment of the problem. 
{We will account for this effect using a kinetic approach which includes simultaneously 
photon conversion and absorption in order to characterize the ALP emission rate}, as we will show in Sec. IV.

Here we limit ourselves to a simple estimation of the conversion probability $P( \gamma_T \rightarrow a)$.
First, we consider the resonant case and we assume that the distance traveled by the photon in the $B$-field is equal to its mean free path
 $\lambda_{\rm mfp}=\Gamma_{\rm{abs}}^{-1}$, { since for larger distances the photon scatterings break the coherence of the oscillations}. For the photon transverse modes the resonance occurs when $m^2_a= \omega^2_{\rm{p}}$, thus $\Delta_{\rm{osc}}^T \equiv \Delta_{a \gamma}^T \ll \Gamma_{\rm{abs}}$ as we see from Fig.~\ref{fig:oscpar}. Therefore in Eq.~(\ref{eq:probconva}) we have that $\sin^2(\Delta_{\rm {osc}}^T \Gamma^{-1}_{\rm {abs}})/(\Delta_{\rm {osc}}^T \Gamma^{-1}_{\rm{abs}})^2 \approx 1$. Then the TP-ALP conversion probability reads
%%%%
\begin{eqnarray}
& & P( \gamma_T \rightarrow a)  \approx  \biggl( \frac{g_{a \gamma} B_T \Gamma_{\rm{abs}}^{-1}}{2} \biggr)^2 \approx 5.34 \times 10^{-22}  \nonumber \\ 
& \times &  \biggl( \frac{g_{a \gamma}}{5 \times 10^{-11} \; \textrm{GeV}^{-1}} \biggr)^2 \biggl( \frac{B_T}{3 \times 10^5 \; \rm{G}} \biggr)^2 \biggl(\frac{ \Gamma_{\rm{abs}}^{-1}}{0.4 \; \textrm{cm}} \biggr)^2 \; . \nonumber \\
\label{eq:st}
\end{eqnarray}
%%%%
Next, we consider the off-resonance conversion probability. For the photon transverse modes, 
when the resonance condition does not apply we can take $m_a\sim 0$, so that $\Delta_{\rm{osc}}^T \approx \Delta_p$. From Fig.~\ref{fig:oscpar}, we see that $\Delta_{\rm p} \gg \Gamma_{\rm{abs}}$, thus in Eq.~(\ref{eq:probconva}) we can approximate $\sin^2(\Delta_{\rm {osc}}^T \Gamma^{-1}_{\rm {abs}}) \approx 1/2$, since there are many photon oscillations within a mean free path $\Gamma_{\rm {abs}}^{-1}$ and we can just consider the average of the oscillatory term. In this case the TP-ALP conversion probability becomes
%%%%
\begin{eqnarray}
& & P( \gamma_T \rightarrow a)  \approx  \frac{1}{2}\biggl( \frac{g_{a \gamma} B_T/2}{\Delta_{\rm{p}}} \biggr)^2
\approx  2.14 \times 10^{-36}
\nonumber \\
& \times &  \biggl( \frac{g_{a \gamma}}{5 \times 10^{-11} \; \textrm{GeV}^{-1}}\biggr)^2 \biggl( \frac{B_T}{3 \times 10^5 \; \rm{G}} \biggr)^2
\nonumber \\
 &\times&  \biggl( \frac{10^{26} \; \textrm{cm}^{-3}}{n_e}\biggr)^2  \biggl( \frac{\omega}{\textrm{keV}} \biggr)^2\; . \nonumber \\
\end{eqnarray}
%%%%
 {Thus, the off-resonance TP-ALP conversion probability is much smaller than the resonant one.
 However, the resonant conversions involve only a small fraction of photons in a given shell of the Sun. Therefore, 
a complete calculation of the ALP flux is necessary to determine which contribution would dominate.}

%..............
\subsection{Longitudinal photons}
%..................

Longitudinal plasmons mix with ALPs in a plasma through an external longitudinal magnetic field. Therefore, we consider 
a uniform external magnetic field along the $z$-direction ${\bf B}_{\rm ext}= B_L  \hat z$. 
If we take  $\omega \simeq \omega_{\rm {p}} \simeq \omega_{a}$, we can linearize the Maxwell's equations for longitudinal modes, obtaining~\cite{Tercas:2018gxv}
%%%%
\begin{equation}
i \partial_z
\begin{pmatrix}
A_L\\
a
\end{pmatrix}
=
H_L
\begin{pmatrix}
A_L\\
a
\end{pmatrix} \; ;
\end{equation}
%%%%
where $A_L$ is the longitudinal plasmon field and $a$ the ALP field and
 the Hamiltonian of the system is 
%%%%
\begin{equation}
H_L=
\begin{pmatrix}
\omega_p & \Delta_{a \gamma}^L \\
\Delta_{a \gamma}^L & \omega_a
\end{pmatrix} \; . 
\label{eq:hlonga}
\end{equation}
%%%%
Then, the LP-ALP conversion probability is~\cite{Mendonca:2019eke}
%%%%
\begin{equation}
P(\gamma_{{L}} \rightarrow a) = (\Delta_{a \gamma}^L z)^2 \frac{\sin^2(\Delta^L_{{\rm osc}} z/2)}{(\Delta^L_{\rm {osc}} z/2)^2} \; ;
\label{eq:lp}
\end{equation}
%%%%
with 
%%%%
\begin{align}
& \Delta_{a \gamma}^L= \frac{g_{a \gamma} B_L}{2}\; ; \\
& \Delta^L_{\rm{osc}}=\sqrt{4 {\Delta_{a \gamma}^L}^2+ ( \omega_a- \omega_{\rm p})^2} \; ,
\end{align}
%%%%
where the numerical value of $\Delta_{a \gamma}^L$ can be calculated as for  $\Delta_{a \gamma}^T$ in Eq.~(\ref{eq:transvparam}).
The LP-ALP conversion presents a  resonance for $\omega=\omega_{\rm p}$, when the ALP dispersion relation (dot-dashed curve) crosses
 the LP one (dashed curve), as shown in Fig.~\ref{fig:dispassione}. 

Concerning the LP absorption, in~\cite{Redondo:2013lna}  it has been shown that for these modes one finds the same expression as for the TP absorption rate.
Therefore, also in this case $\Gamma_{\rm abs} \gg \Delta_{a\gamma}^L$.
Concerning the resonant conversion probability, one finds  the same numerical expression of Eq.~(\ref{eq:st}) for the TP.
%%%%

%%%%%%%%%%%%%%%%%%%%%%
\section{ALP Production rate}
\label{sec:ALP_rates}
%%%%%%%%%%%%%%%%%%%%%%
\label{sec:rate}

As we have seen in the previous Section, the photon absorption rate $\Gamma_{\rm{abs}}$ in the Sun is not negligible with respect to the others oscillation parameters. Thus, TP-ALP and LP-ALP oscillations are interrupted by collisions 
and we have to face the problem of treating simultaneously oscillations and collisions.
A suitable formalism is provided by  the \textit{kinetic approach} developed for  relativistic mixed neutrinos in the presence of collisions~\cite{Sigl:1992fn}.
This formalism   has been applied to different mixing problems, such as the mixing of photons with 
hidden photons (HP) in the Sun~\cite{Redondo:2013lna}, which we closely follow in our derivation. Details are given in Appendix B.
In Appendix C, we show that this approach is equivalent to the thermal field theory formalism.

To begin, we 
present a completely general formalism, starting from two bosonic fields $A$ and $S$, which evolve according to the linearized equation of motion
%%%%
\begin{equation}
i \partial _t 
\begin{pmatrix}
A \\
S
\end{pmatrix}
=
\begin{pmatrix}
\omega_A & \mu \\
\mu & \omega_S
\end{pmatrix}
\begin{pmatrix}
A\\
S
\end{pmatrix}\; ;
\label{eq:lin}
\end{equation}
%%%%
where $\omega_A$ is the energy associated with the field $A$, $\omega_S$ the one associated with the field $S$, 
and $\mu$ is a mixing term which we assume to be small with respect to the diagonal terms.
The Hamiltonian in Eq.~(\ref{eq:lin}) can be written as
%%%%
\begin{equation}
H=
\begin{pmatrix}
\omega_A & \mu \\
\mu & \omega_S 
\end{pmatrix}
= \frac{\omega_A+\omega_S}{2} I +
\begin{pmatrix}
\frac{1}{2}\Delta \omega & \mu \\
\mu & -\frac{1}{2} \Delta \omega
\end{pmatrix} \; ;
\label{eq:diag}
\end{equation}
%%%%
where $\Delta \omega= \omega_A-\omega_S$. For such a system, we can define the oscillation frequency
%%%%
\begin{equation}
\Delta_{\rm{osc}}= \sqrt{4 \mu^2+ \Delta \omega ^2}\; .
\label{eq:oscfrequency}
\end{equation}
%%%%
We consider the case in which collisions occur for the $A$ quanta (i.e. the photons in our case). We assume that the field $A$ interacts with the medium, namely with the solar plasma, which can absorb a quantum with rate $\Gamma_{\rm{abs}}$ and produce one with rate $\Gamma_{\rm{prod}}$.
The equations of motion for such a system, described by the density matrix $\rho$, are the Liouville equations~\cite{Sigl:1992fn}
%%%%
\begin{equation}
\dot{\rho}=-i[H, \rho]+\frac{1}{2} \{G_{\rm{prod}}, I+ \rho \}-\frac{1}{2} \{G_{\rm {abs}}, \rho \} \; ;
\label{eq:liouvillea}
\end{equation}
%%%%
where
%%%%
\begin{equation}
\begin{split}
G_{\rm {prod}}=
\begin{pmatrix}
\Gamma_{\rm{prod}} & 0 \\
0 & 0
\end{pmatrix} \; ; \\
G_{\rm{abs}}=
\begin{pmatrix}
\Gamma_{\rm{abs}} & 0 \\
0 & 0
\end{pmatrix} \; \; . 
\end{split}
\end{equation}
%%%%
We remind the reader that the diagonal components of the density matrix contain the occupation numbers of $A$ and $S$ quanta, while 
the off-diagonal components take into account the coherence between these two states.
Note that in Eq.~(\ref{eq:liouvillea}) the commutator on right-hand-side describes the dynamical evolution of the system while the anticommutators correspond to the collisional terms.
In thermal equilibrium $\Gamma_{\rm{prod}}= e^{-\omega(k)/T} \Gamma_{\rm{abs}}$ and the $S$  type particles are not excited, while we assume that the $A$ type particles obey the Bose-Einstein statistics $f_{\rm{BE}}=(e^{\omega(k)/T}-1)^{-1}$, where $\omega(k)$ is the photon energy. A non-equilibrium situation is described with a small deviation $\delta \rho$ from the thermal equilibrium state $\rho_{\rm{eq}}$.
In this limit one finds a steady state solution for the $S$ quanta production rate 
%%%%
\begin{equation}
\Gamma_S^{\rm{prod}} \equiv \dot{n}_S=  \frac{\Gamma \mu^2}{(\omega_A-\omega_S)^2 + \Gamma^2/4} \frac{1}{e^{\omega(k)/T}-1} \; ,
\label{eq:rategeneralea}
\end{equation}
%%%%
 with $\Gamma=(1-e^{-\omega(k)/T})\Gamma_{\rm{abs}}$, i.e. the total collisional rate. For simplicity of notation we will denote $\omega(k) \equiv \omega$, implying its dependence on the photon moment $k$.
From Eq.~(\ref{eq:rategeneralea}) we obtain that the $A-S$ mixing process is \textit{resonant}, i.e. it is maximal for $\omega_A=\omega_S$. 
The result in Eq.~(\ref{eq:rategeneralea}) is completely general and it is valid both at {the resonance} and off-resonance, 
since it was obtained under the only assumption that the mixing term $\mu$ is small relatively to the diagonal terms $\sim \Delta \omega$. 
This condition always applies in the solar plasma, {for both TP and LP}. 
Thus, we can adopt Eq.~(\ref{eq:rategeneralea}) to describe the TP-ALP and LP-ALP conversion rates.

%%%%%%%%%%%%%%%%%%
\subsection{Photon transverse modes}
%%%%%%%%%%%%%%%%%%%
The Hamiltonian for TP-ALP system 
in Eq.~(\ref{eq:htransa}) can be written  (up to a term proportional to the identity matrix) as in Eq.~(\ref{eq:diag}) 
\begin{equation}
H_T=
\begin{pmatrix}
 q/2 & \Delta_{a \gamma}^T \\
\Delta_{a \gamma}^T & - q/2
\end{pmatrix}
\; ;
\end{equation}
%%%%
where here $q=(\omega^2_{\rm{p}}-m^2_a)/2 \omega$. Thus, with the substitutions $\mu \rightarrow \Delta_{a \gamma}$ and $\Delta \omega \rightarrow q$, 
from Eq.~(\ref{eq:rategeneralea})
we obtain  the TP-ALP conversion rate
%%%%
\begin{equation}
\Gamma_a^{\rm{prod}}= \biggl[\frac{\Gamma {\Delta_{a \gamma}^T}^2}{\bigl(\Delta_{\rm p}  - \Delta_{\rm a} \bigr)^2 + \Gamma^2/4} \biggr] \frac{1}{e^{\omega/T}-1} \; .
\label{eq:ratetrasverso}
\end{equation}
%%%%
The expression in Eq.~(\ref{eq:ratetrasverso}) is valid both on resonance and off-resonance and we will use it to estimate the ALP flux expected at Earth arising from these conversion processes.

%%%%%%%%%%%%%
\subsection{Photon longitudinal modes}
%%%%%%%%%%%%%%%%%%%%
The Hamiltonian for LP-ALP system 
in Eq.~(\ref{eq:hlonga}) can be written  as in Eq.~(\ref{eq:diag}) 
 %%%%
\begin{equation}
H_L=
\begin{pmatrix}
 \Delta/2 \omega & \Delta_{a \gamma}^L \\
\Delta_{a \gamma}^L & -\Delta/2 \omega
\end{pmatrix} 
\; ;
\end{equation}
%%%%
where in this case $\Delta \omega=\omega_{\rm p}- \omega_a$. If we insert this expression of $\Delta \omega$ and we replace $\mu \rightarrow \Delta_{a \gamma}^L$ in Eq.~(\ref{eq:rategeneralea}) we obtain the LP-ALP conversion rate
%%%%
\begin{equation}
\Gamma_a^{\rm{prod}}= \biggl[ \frac{\Gamma {\Delta_{a \gamma}^L}^2}{(\omega_{\rm p}-\omega_a)^2 + \Gamma^2/4} \biggr] \frac{1}{e^{\omega/T}-1} \; .
\label{eq:ratelongitudinale}
\end{equation}
%%%%
Contrarily to the photon transverse modes, the expression in Eq.~(\ref{eq:ratelongitudinale}) is valid only on resonance, 
since it is based on the evolution of on-shell LPs, i.e. it is obtained assuming that $\omega \sim \omega_{\rm p} \sim \omega_a$ and it is not applicable for $\omega$ very far from $\omega_{\rm p}$.
On the other side, a general result free of these limitations 
 has been recently obtained in~\cite{Caputo:2020quz,OHare:2020wum} from a thermal field theory calculation (Cfr. Appendix C).
 We checked that on resonance this latter result [see Eq.~(\ref{eq:gammalth})] agrees with our previous one.

%%%%%%%%%%%%%%%%%%%%%%
\section{Solar ALP fluxes }
\label{sec:ALP_fluxes}
%%%%%%%%%%%%%%%%%%%%%%

{The solar ALP flux on Earth is given by}~\cite{Andriamonje:2007ew}
%%%%
\begin{equation}
\frac{dN_a}{dt}= \frac{g}{4 \pi D_\odot^2} \int d^3{{\bf r}} \frac{d^3 {{\bf k}}}{(2 \pi)^3} \Gamma_a^{\rm{prod}} \; ;
\label{eq:number}
\end{equation}
%%%%
where $D_\odot =1.49 \times 10^{11} \; \rm{m}$ is the Earth-Sun distance, $\Gamma_a^{\rm{prod}}$ is the ALP production rate expressed by 
Eq.~(\ref{eq:rategeneralea}), the factor $g$ is the number of the photon polarization states ($g=1$ for LP and $g=2$ for TP), 
and the integral is performed over the photon
momenta ${\bf k}$ and over the solar volume.
From Eq.~(\ref{eq:number}) we recover the differential ALP {spectrum} expected at Earth
%%%%
\begin{equation}
\frac{d \Phi_a}{d \omega}= \frac{g}{(2 \pi)^3 D_\odot^2}\int_0^{R_\odot} dr r^2 \int d \Omega_{{\bf k}} \,\ \omega^{2}
 \Gamma_a^{\rm{prod}} \; ,
\label{eq:diff}
\end{equation}
%%%%
where we assumed relativistic states $\omega \approx k$ and $ \Omega_{{\bf k}}$ is the solid angle around the direction of photon 
momentum ${\bf k}$.
 In the following we will perform the radial integral over the SSM  AGSS09 \cite{Serenelli:2009yc}.
We now focus on the estimation of the ALP flux at Earth from different conversion processes in the solar magnetic fields for TP and LP modes.

%%%%%%%%%%%%%%%
\subsection{Flux from TP-ALP conversions}
%%%%%%%%%%%%%%%%%%%%%%

The TP-ALP conversion process is dominated by the resonance, where the TP-ALP conversion probability [Eq.~(\ref{eq:probconva})] is maximal.
Due to the    extremely peaked nature of the resonant condition, we can approximate the ALP production rate [Eq.~(\ref{eq:ratetrasverso})] with a delta function
%%%%
\begin{eqnarray}
\Gamma_{a,T}^{\rm{prod}}& \approx& 2 \pi  {\Delta_{a \gamma}^T}^2 \delta \biggl(\Delta_{\rm p}-\Delta_{\rm a} \biggr) \frac{1}{e^{\omega/T}-1}
\nonumber \\
& \approx & \frac{\pi}{2}(g_{a \gamma} B_T)^2 \delta \biggl( \frac{\omega^2_{\rm{p}}-m^2_a}{2 \omega} \biggr)  \frac{1}{e^{\omega/T}-1}\; .
\label{eq:gammat}
\end{eqnarray}
%%%%
If we insert the last expression in Eq.~(\ref{eq:diff}), we note that the integration over the solar volume gives
%%%%
\begin{equation}
\int dr \delta (q) \approx {\left| \frac{d q}{dr} \right|^{-1}_{\rm{res}}}= 2 \omega {\left| \frac{d \omega_{\rm{p}}^2}{ dr} \right|^{-1}_{\rm{res}}}\; .
\label{eq:integ}
\end{equation}
%%%%
{To evaluate the above expression, we model the electron density in the region $r \lesssim 0.8 \; R_\odot$
	with a simple exponential form 
	%%%%
	\begin{equation} 
		n_e= n_e^0 e^{-r/R_e} \,\ .
	\end{equation} 
	%%%%
The best fitting parameters for the solar model AGSS09 are 
	%%%%
	\begin{align}
		& R_e=R_\odot /9.89 \; \\
		& n_e^0= 1.11 \times 10^{26} \; \textrm{cm}^{-3} \;. 
\end{align}
We thus find
}
% Finally 
\begin{equation}
{\left| \frac{d\omega_{\rm p}^2}{dr} \right|^{-1}_{\rm res}} = \frac{1}{m_a^2} {\left| \frac{d \ln n_e}{dr} \right|^{-1}_{\rm res}}= \frac{1}{m_a^2} R_e \; .
\label{eq:result}
\end{equation}
%%%%
%where $R_e$ is a parameter introduced assuming an exponential model for the electron density in the Sun
%in the region $r \lesssim 0.8 \; R_\odot$
%%%%%
%\begin{equation} 
%n_e= n_e^0 e^{-r/R_e} \,\ ,
%\end{equation} 
%%%%%
% with
%%%%%
%\begin{align}
%& R_e=R_\odot /9.89 \; \\
%& n_e^0= 1.11 \times 10^{26} \; \textrm{cm}^{-3} \;. 
%\end{align}
%%%%%

Furthermore, we notice that in the case of resonance $\Delta_{\rm osc}^T \equiv \Delta_{a \gamma}^T \ll \Gamma_{\rm abs}$
(see Fig.~\ref{fig:oscpar}). Therefore, during the resonance the photons are  continuously
re-scattered  such that information about their polarization is lost. 
The photon trajectories can form any angle with the magnetic field ${\bf B}$. 
Since the photon trajectories are not straight, this  angle is not correlated with the magnetic field direction and the photon polarization. 
Therefore, we have to perform a local angular average in the resonance shell before performing the  integral in
$d\Omega_{\bf k}$ of Eq.~(\ref{eq:diff}).
For a generic photon polarization, the $B_T$ strength entering the conversion probability is 
%...............
\begin{equation}
B_T=|{\bf{B}({\bf z})\cdot {\hat \epsilon}}|=|{\bf B}({\bf z}) \sin \vartheta({\bf z}) \cos \varphi| \,\ ,
\end{equation}
%.................
where ${\bf z}$ is the position vector of the resonance region in a particular direction ${\hat z}$,
${\hat \epsilon}$ is the photon polarization vector ($|{\hat \epsilon}|=1$, 
${\hat \epsilon} \times {\hat z}=0$), $\vartheta$ is the angle between the magnetic field 
 ${\bf B}({\bf z})$ and the 
photon propagation direction ${\hat x}$ and $\varphi$ the angle between
${\bf B}_T$ (the component of the magnetic field perpendicular to ${\bf z}$) and
${\hat \epsilon}$.
We define
%...............
\begin{equation}
\langle B_T^2 \rangle = |{\bf B}|^2 \int \frac{d \varphi}{2 \pi} \frac{d \Omega_{\vartheta}}{4 \pi}
\sin^2 \vartheta \cos^2\varphi = \frac{1}{3}  |{\bf B}|^2 \,\ .
\label{eq:magnetic average}
\end{equation}
%..................
Therefore, in Eq.~(\ref{eq:gammat}) we should substitute $B_T^2 \to \langle B_T^2 \rangle =  |{\bf B}|^2/3 $.

Finally, the ALP flux spectrum from resonant conversions in the solar magnetic field is obtained inserting
 Eqs.~(\ref{eq:integ})--(\ref{eq:result})--(\ref{eq:magnetic average}) into Eq.~(\ref{eq:diff}) obtaining
%%%%
\begin{equation}
\frac{d \Phi_{a,T}}{d\omega}= \frac{1}{ 3 \pi D_{\odot}^2} \biggl( \frac{g_{a\gamma} |{\bf B}(r_{\rm res})|}{m_a} \biggr)^2r_{\rm res}^2 R_e 
 \frac{\omega^3}{e^{\frac{\omega}{T_{\rm res}}}-1} \; ; \label{eq:resonantflux}
\end{equation}
%%%%
where $r_{\rm{res}}$ is the position in the Sun where the resonance condition occurs for the fixed value of $m_a$ and $T_{\rm{res}}$ is the temperature at the same position.

Let us first consider the ALP spectrum for the resonance in the tachocline at $r\sim 0.7 \; R_\odot$, where $m_a \sim 10 \; \eV$ and $B \sim 3 \times 10^5 \; \rm{G}$, where   the homogeneous $B$-field has its peak, as shown in Fig.~\ref{fig:bfields}. 
An analytic approximation to the solar ALP  spectrum is provided by a fit with the three-parameter function~\cite{Andriamonje:2007ew}
%%%%
\begin{equation}
\frac{ {\rm d} \Phi_{a,T}}{ {\rm d} \omega}=g_{10}^2 C \biggl( \frac{\omega}{\omega_0} \biggr)^{\alpha} e^{-(\alpha + 1) \frac{\omega}{\omega_0}} \; ;
\label{eq:fitres}
\end{equation}
%%%%
where $C$ is a normalization constant, the energy $\omega$ is expressed in $\keV$ and $\omega_0$ is an energy scale with the property $\omega_0= \langle \omega \rangle$. 
Numerical values of $C$, $\alpha$ and $\omega_0$ are shown in Table~\ref{tab:parameters}.
%%%%%%%%%%%%%%
\begin{table}[!t]
  \caption{Parameters of the solar ALP spectrum for different values of 
$m_a$.}
\begin{center}
\begin{tabular}{lclclc|c}
\hline
$m_a (\textrm{eV}) $ & $C$ & $\omega_0$ (keV) & $\alpha$  \\
\hline
\hline
10 & 9.4 $\times 10^{11}$ & 0.61 & 2.46  \\
130 & 1.36 $\times 10^{15}$& 2.80 & 2.47 \\
0 & 8.3 $\times 10^{10}$ & 3.15 & 3.16 \\
\hline
\end{tabular}
\label{tab:parameters}
\end{center}
\end{table}
%%%%%%%%%%%%%%%%%%%%%%%%%

We can obtain the total ALP flux $\Phi_{a,T}$ from resonant conversions in the solar magnetic fields expected at Earth by integrating Eq.~(\ref{eq:resonantflux}) [or equivantely Eq.~(\ref{eq:fitres})] over the energies $\omega$.
Using Eq.~(\ref{eq:fitres}) and the fit parameters in Table \ref{tab:parameters} we obtain the following flux quantities for $m_a= 10 \; \eV$
%%%%
\begin{align}
& \Phi_{a,T}= 2.48 \times 10^{10} \; g_{10}^2 \; \cm^{-2} \; \rm{s}^{-1}  \; ; \\
& \langle \omega \rangle = 0.6 \; \keV  \;  ;\\
& L_{a,T}=1.51  \times 10^{-5} \; g_{10}^2 \; L_\odot  \; ; 
\label{eq:res10eV}
\end{align}
%%%%
where $L_{a,T}$ is the ALP luminosity, $L_\odot=3.8418 \times 10^{33} \; \rm{erg} \; \rm{s}^{-1}$ is the Sun luminosity and $g_{10}=g_{a \gamma}/10^{-10} \; \GeV^{-1}$.

We now focus  on the resonant production at $r \sim 0.25 \; R_\odot$, i.e. the one with $m_a=130 \; \eV$. 
Here we present results obtained assuming $B=3 \times 10^{7} \; \rm{G}$. We find the following flux quantities
%%%%
\begin{align}
& \Phi_{a,T}= 1.63 \times 10^{14} \; g_{10}^2 \; \cm^{-2} \; \rm{s}^{-1} \; \keV^{-1} \; ; \\
& \langle \omega \rangle = 2.76 \; \keV \; ; \\
& L_{a,T}=0.2 \; g_{10}^2 \; L_\odot \; . \label{eq:lumgrande}
\end{align}
%%%%

Finally, if we are far from resonance we can assume $m_a \approx 0$. In this case $\Delta_{\rm{osc}}^T \approx \Delta_{\rm{p}} \gg \Gamma_{\rm{abs}}$, as shown in Fig.~\ref{fig:oscpar}. Thus, the rate in Eq.~(\ref{eq:ratetrasverso}) reduces to
%%%%
\begin{equation}
\Gamma_{a,T}^{\rm{prod}}\simeq \Gamma_{\rm{abs}} (1- e^{-\omega/T}) \frac{{\Delta^T_{a \gamma}}^2}{(\omega_{\rm p}^2/2 \omega)^2} \frac{1}{e^{\omega/T}-1} \; .
\label{eq:offres}
\end{equation}
%%%%
{
If we insert Eq.~(\ref{eq:offres}) in Eq.~(\ref{eq:diff}) and we integrate over the solar profile we obtain the ALP flux spectrum at Earth from off-resonant production.
In this case  the direction $\vartheta$ between the field $\textbf{B}$ and the photon direction of propagation does not change during 
the conversions, since many oscillations occur into a single photon mean free path. However, the azimuthal angle $\varphi$ between the transverse field $B_T$ and the photon polarization $\hat{\epsilon}$ would  change. Thus, we perform an average over $\varphi$ 
before the integral over $d \Omega_{ \bf k}$ in Eq.~(\ref{eq:diff}), i.e. 
%..............
\begin{equation}
\int d\Omega_{\bf k} \int_0^{2 \pi} \frac{d \varphi}{2 \pi} |{\bf B}|^2 \sin^2 \vartheta({\bf x}) \cos^2 \varphi = \pi \frac{4}{3} |{\bf B}|^2 \; . 
\label{eq:angint}
\end{equation}
%........
}
 The dominant contribution for the off-resonant flux comes from the radiative zone, since here the $B$-field amplitude reaches the highest value. 
Taking the peak value
$B= 3 \times 10^{7} \; \rm{G}$
 we find the following flux quantities for non-resonant ALP spectrum flux
%%%%
\begin{align}
& \Phi_{a,T}=5.2 \times 10^{9} g_{10}^2 \; \cm^{-2} \; \rm{s}^{-1} \; ; \\
& \langle \omega \rangle= 3.15 \; \keV \; ; \\
& L_{a,T}= 1.92 \times 10^{-8}\; g_{10}^2 \; L_\odot \; .
\end{align}
%%%%%%%%%%%%%%%%%%%%%

\begin{figure}[t]
\centering
\includegraphics[width=\columnwidth]{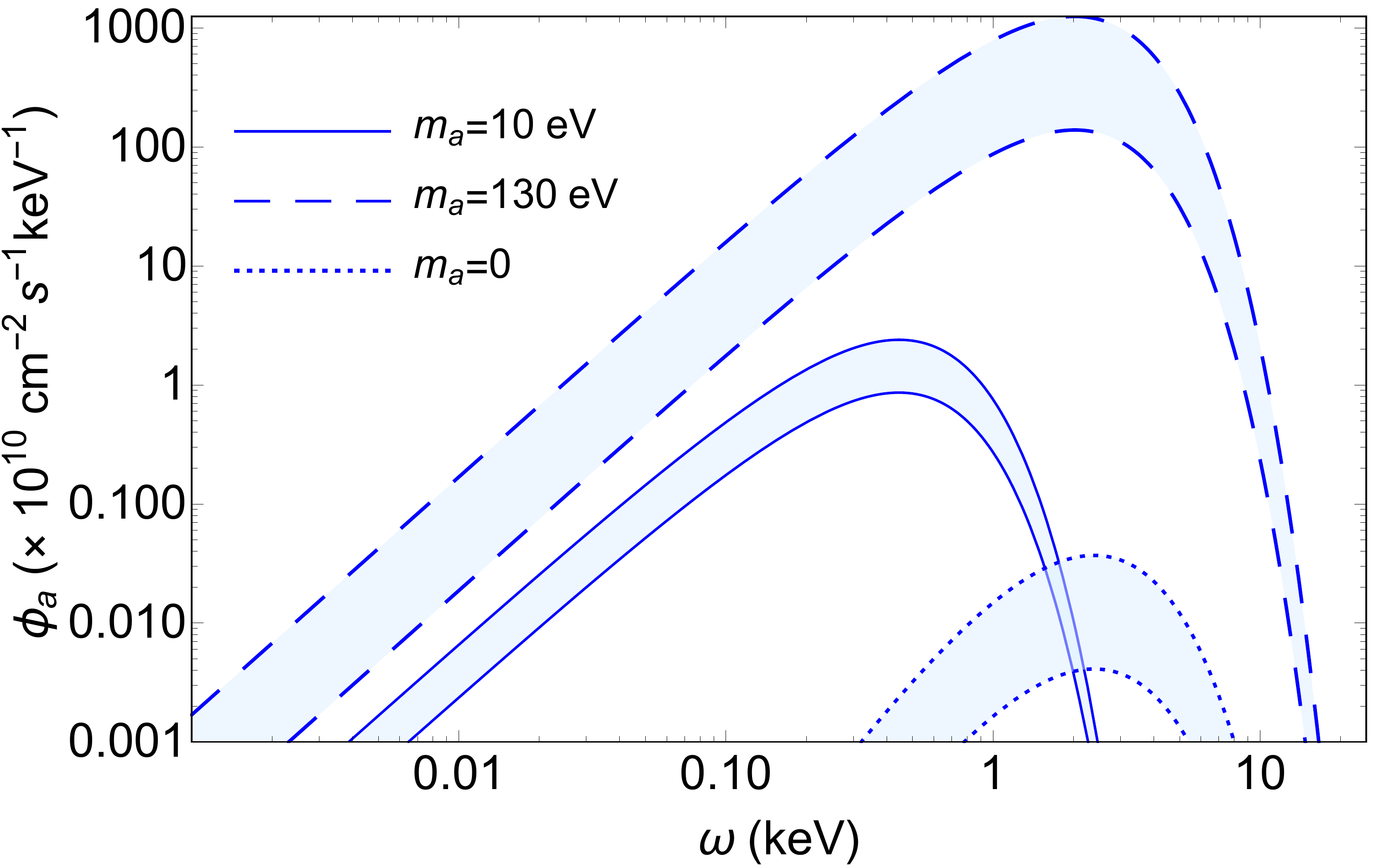}
\caption{Comparison between the fluxes from TP-ALP conversions in solar magnetic fields
for resonant conversions in the tachocline for $m_a=10$~eV  (continuous curve), in the radiative zone for $m_a=130$~eV (dashed curve)
and non-resonant case for $m_a=0$ (dotted curve). The bands width reflects the uncertainty related to the B-field models [See Fig.~\ref{fig:bfields}].
 The largest contribution is given by the resonant production. 
 We assume $g_{a \gamma}= 5 \times 10^{-11} \; \GeV^{-1}$.}
\label{fig:tuttiris}
\end{figure}

We report in Table~\ref{tab:parameters} the fitting parameters of the energy spectrum of Eq.~(\ref{eq:fitres}) for the three cases we considered.
In Fig.~\ref{fig:tuttiris}, we compare the fluxes from TP-ALP conversions in solar magnetic fields. As expected,  we see that the non-resonant contribution is always subdominant with respect to the resonant ones. The band width in each flux contributions represents the uncertainty coming from the magnetic field models, as shown in Fig.~\ref{fig:bfields}.
%%%%%%%%%%%%%%%%%%%%
%%%%%%%%%%%%%%%%%%%%%%%%%%%%

%%%%%%%%%%%%%%%
\subsection{Flux from LP-ALP conversions}
%%%%%%%%%%%%%%%%%%%%%%

\begin{figure}[t]
\centering
\includegraphics[width=\columnwidth]{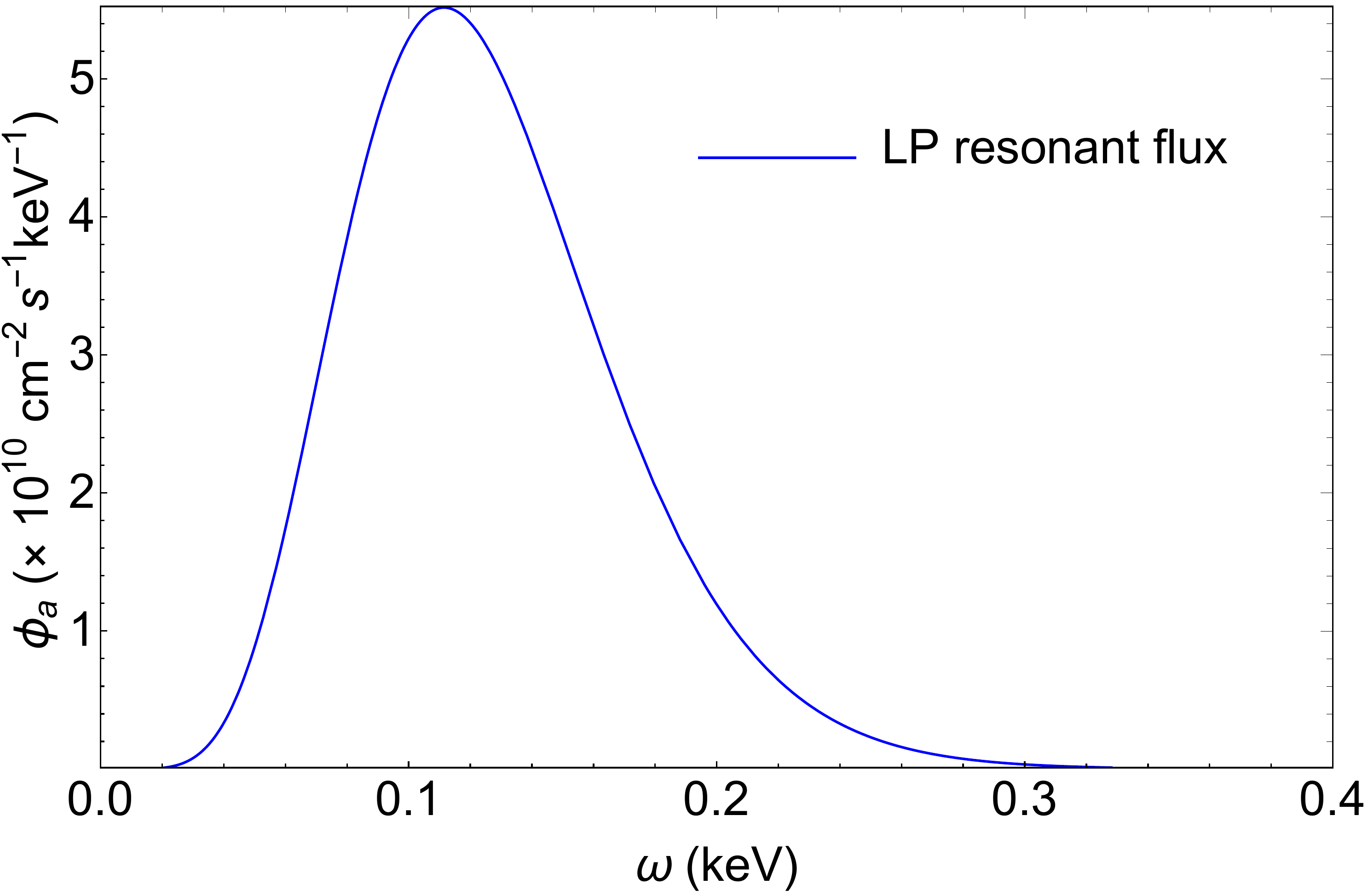}
\caption{ALP flux expected at Earth from LP-ALP conversions in the Solar magnetic field of the radiative zone ($B=3 \times 10^{7} \; \rm{G}$). The flux was computed assuming $g_{a \gamma}= 5 \times 10^{-11} \; \GeV^{-1}$.}
\label{fig:lpflux}
\end{figure}

The ALP production rate for LP-ALP conversions in the solar magnetic fields is given in  Eq.~(\ref{eq:ratelongitudinale}).
 Since the largest contribution arises from conversions in the radiative zone, we will present results just for the largest value of the $B$-field in this region, i.e. $B = 3 \times 10^7 \; \rm{G}$.
Also in the case of LP-ALP conversions the process is extremely peaked around the resonance, thus we can approximate $\Gamma_{a,L}^{\rm{prod}}$ with a delta function
%%%%
\begin{equation}
\Gamma_{a,L}^{\rm{prod}} \approx 2 \pi {\Delta_{a \gamma}^L}^2 \delta( \omega_{\rm p}- \omega_a) \frac{1}{e^{\omega/T}-1} \;.
\end{equation}
%%%%
In the  case of LP there is just one projection of the magnetic field which is longitudinal to the photon propagation direction, i.e. $B_L= 
|{\bf B}  \cos \vartheta|$. Then, in the resonant shell we should average
	\begin{equation}
		\langle B_L^2 \rangle =
		\int \frac{d \Omega_{\vartheta}}{4 \pi} |{\bf B}|^2 \cos^2 \vartheta 
		= \frac{1}{3} |{\bf B}|^2 \; .
	\end{equation}
{If we insert this expression in Eq.~(\ref{eq:diff}) we obtain the ALP flux from LP-ALP conversion in the Sun
%%%%
\begin{equation}
\begin{split}
\frac{d \Phi_{a,L}}{d \omega}
=\frac{1}{12 \pi D_\odot^2}r_{\rm res}^2 \frac{\omega^2 g_{a \gamma}^2 |{\bf B}(r_{\rm res})|^2}{e^{\omega/T}-1} \frac{1}{| \omega^\prime(r_{\rm res})|} \,\ ,
\label{eq:flong}
\end{split}
\end{equation}
%%%%
}
where $r_{\rm res}$ is the position in the Sun where the resonance $\omega=\omega_{\rm{p}}$ occurs, 
$|\omega^\prime(r_{\rm res})|= |d\omega_p/dr|$ computed at $r=r_{\rm res}$ and we have 
denoted with $\omega \approx \omega_a \approx \omega_{p}$ the ALP and photon energies. 
The result in Eq.~(\ref{eq:flong})
agrees with the one recently obtained  in Ref.~\cite{OHare:2020wum}. 
Using Eq.~(\ref{eq:result}), the derivative in Eq.~(\ref{eq:flong}) can be expressed as 
%%%%
\begin{equation}
\biggl|\frac{d \omega_p}{dr}\biggr|^{-1}_{{\rm res}}= \frac{2R_e}{\omega_p}\; .
\end{equation}
%%%%
%where we used Eq.~(\ref{eq:result}). 
Adopting the Standard Solar Model AGSS09 \cite{Serenelli:2009yc} to compute the plasma frequencies we obtain the ALP flux spectrum shown in Fig.~\ref{fig:lpflux}.
The flux shows a peak at $\omega \sim 0.12\; \keV$. The flux quantities are found to be
%%%%
\begin{align}
& \Phi_{a,L}=2.18 \times 10^{10} \; g_{10}^2  \; \cm^{-2} \; \rm{s}^{-1} \; ; \label{eq:fi} \\
& \langle \omega \rangle= 0.13 \; \keV\; ; \label{eq:enm} \\
& L_{a,L}= 3.34 \times 10^{-6} \; g_{10}^2  \; L_\odot \,\ . \label{eq:lu}
\end{align}
%%%%

%%%%%%%%%%%%%%%%%%%%%%%
\section{Energy-loss bounds}
\label{sec:Energy-loss}
\begin{figure}[t]
\centering
\includegraphics[width=\columnwidth]{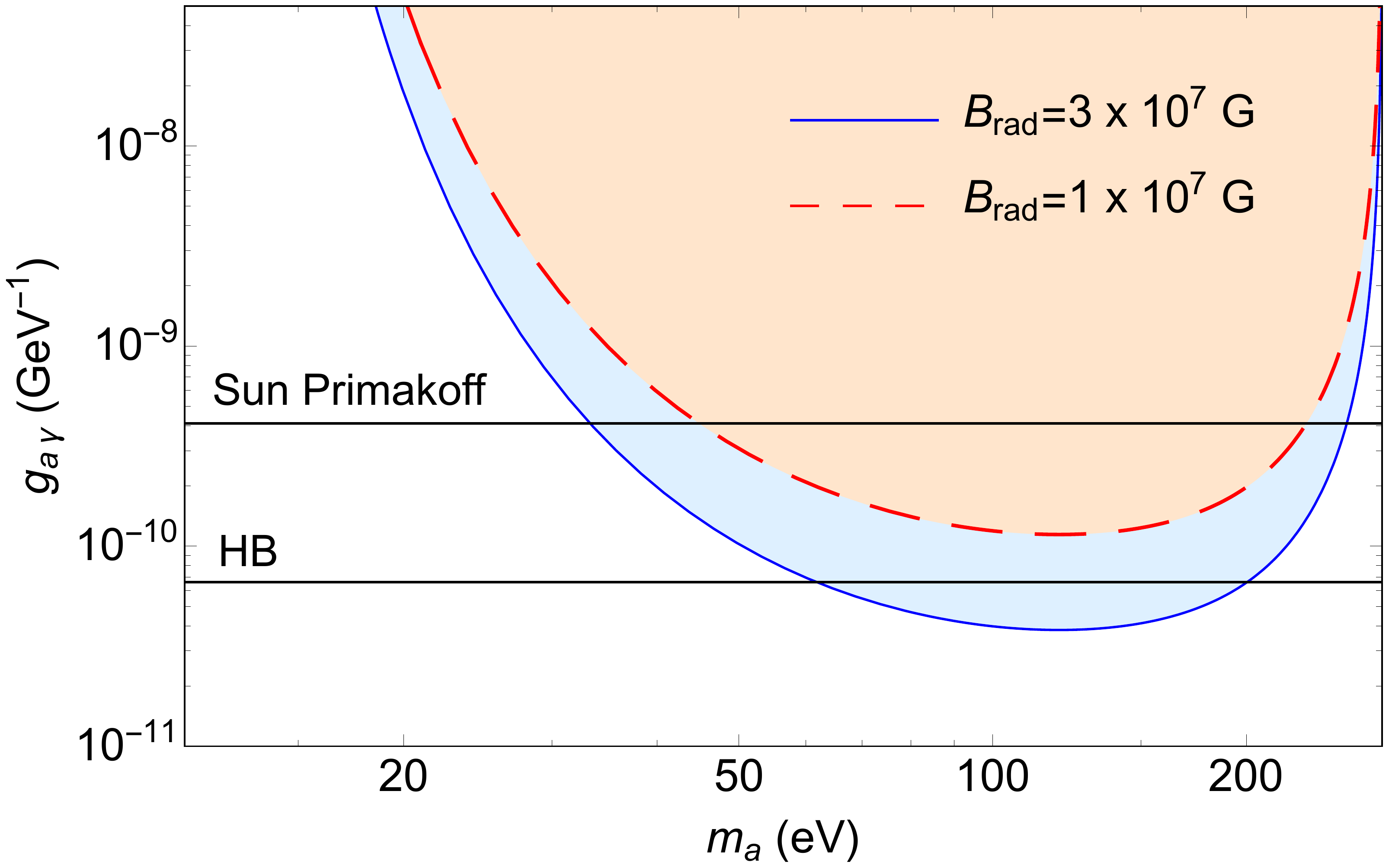}
\caption{The blue region is the excluded one in the parameter space $(m_a, g_{a \gamma})$ by resonant processes in the radiative zone of the Sun assuming a field with amplitude $B_{\rm{rad}}= 3\times 10^7 \; \rm{G}$. The orange region is the excluded one by the same process, assuming a field with amplitude $B_{\rm{rad}}=1 \times 10^7 \; \rm{G}$. The horizontal black lines represent the bound set by the Helium Burning stars (HB) and by Primakoff ALP emission in the Sun.}
\label{fig:boundnuovo}
\end{figure}

%%%%%%%%%%%%%%%%%%%%%%%%%%%
We now discuss the phenomenological consequences of these solar ALP fluxes.
 We start considering the possibility to place  a new bound on ALP-photon coupling $g_{a \gamma}$ based on the ALP emissivity from
 photon conversions in
 $B$-fields.
 On the basis of the energy-loss argument one can set a bound on the coupling $g_{a \gamma}$ imposing the condition~\cite{Vinyoles:2015aba}
%%%%
\begin{equation}
L_a \lesssim 0.03 \; L_\odot \; ,
\label{eq:bound}
\end{equation}
%%%%
which is obtained from the combination of helioseismology (sound speed, surface helium and convective radius) and solar neutrino observations.
Assuming the usual ALP emission by Primakoff process, whose estimated luminosity is $L_a = 1.8 \times
10^{-3} \; g_{10}^2 \; L_\odot $~\cite{Andriamonje:2007ew},
 the quoted bound is $g_{\rm a\gamma}\lesssim 4.1 \times 10^{-10}$~GeV$^{-1}$. 
Now we consider the case of the ALP flux from resonant conversions 
 for masses $m_a \sim {\mathcal O} (100) \; \eV$. Using  the luminosity associated with this flux
we can  set the upper limit on $g_{a \gamma}$ as shown in Fig.~\ref{fig:boundnuovo}
where the blue region is the excluded one in the parameter space $(m_a, g_{a \gamma})$ by resonant processes in the radiative zone of the Sun assuming a field with amplitude $B_{\rm{rad}}= 3\times 10^7 \; \rm{G}$. The orange region is the excluded one by the same process, assuming a field with amplitude $B_{\rm{rad}}=1 \times 10^7 \; \rm{G}$. 
In particular,
%%%%
\begin{eqnarray}
& & g_{a \gamma} \lesssim  (3.8-11.2) \times 10^{-11} \; \GeV^{-1} \nonumber \\
& &\; \textrm{for} \; 100 \; \eV \lesssim m_a \lesssim 140  \eV.
\label{eq:nostro}
\end{eqnarray} 
%%%%
This new bound is even more stringent than the 
constraint derived from Helium burning stars in GCs, $g_{a \gamma}< 6.6 \times 10^{-11} \; \GeV^{-1}$~\cite{Ayala:2014pea}. 
For ALP masses outside the previous  range, the bound
worsens, as shown in Fig.~\ref{fig:boundnuovo}.
For instance, for $m_a \sim 30 \; \eV$ we obtain a limit
comparable with the one from the Primakoff process. 
For conversions of TP in the case of $m_a \sim 10$ eV and $m_a=0$ and for  the LP conversions, the associated ALP luminosity is much smaller than the Primakoff one, as results from Eq.~(\ref{eq:res10eV}) and
Eq.~(\ref{eq:lu}), respectively. Therefore, the contribution to the energy-loss is subleading with respect to the Primakoff process.

%%%%

%%%%%%%%%%%%%%%%%%%%%%%
\section{Detection perspectives}
\label{sec:detection}
%%%%%%%%%%%%%%%%%%%%%%%%%%%

\begin{figure}[t]
\centering
\includegraphics[width=\columnwidth]{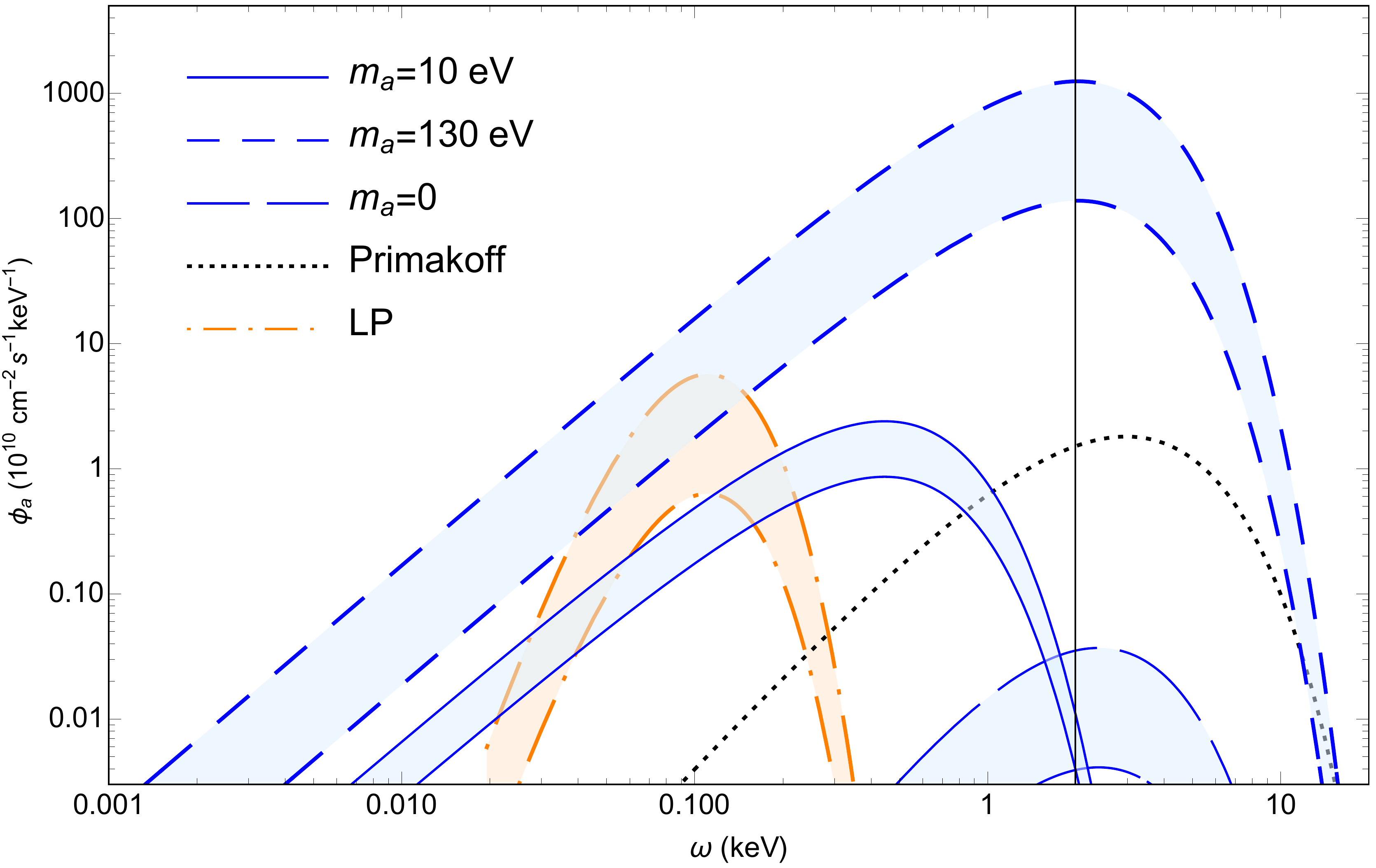}
\caption{Contributions from all ALP fluxes from the Sun. The dotted  line represents the Primakoff flux. The continuous and the short-dashed blue lines are the flux from the resonant conversion in the solar magnetic fields for an axion mass $m_a=10 \; \eV$ (in the tachocline) and $m_a=130 \; \eV$ (in the radiative zone), respectively. 
The long-dashed curve represents the ALP flux from non-resonant conversions of TP modes.
The dot-dashed  line is the flux from the LP-ALP conversions. The vertical line is the CAST energy threshold. The bands width represents the uncertainty associated with the magnetic field model, as shown in Fig.~\ref{fig:bfields}.}
\label{fig:fluscomparison}
\end{figure}

In order to assess the possibility to detect the solar ALP fluxes from conversions in $B$-fields we show in Fig.~\ref{fig:fluscomparison}
the different ALP fluxes from conversions in $B$-fields  and 
we compare them with the Primakoff flux (dotted curve, see, e.g.~\cite{Calore:2020tjw} for a recent calculation), which represents a benchmark
for experimental searches on solar ALPs.
Starting from the ALP flux from TP conversions, we see that 
for $m_a \sim 10 \; \eV$ the ALP flux (continuous curve) is  peaked at energies  below the CAST threshold~\cite{Andriamonje:2007ew} ($\omega < 2 \; \keV$),
shown as vertical line in the Figure.
There are plans to lower the threshold in the sub-keV region in the future helioscope IAXO~\cite{Armengaud:2019uso}. 
However, masses larger than  $m_a \sim 1 \; \eV$ are not accessible even to this experiment.
For these large masses, the coherence of ALP-photon conversions in the magnetic field of the helioscopes is lost and consequently the sensitivity is rapidly reduced.
In principle, there are ideas for a new class of  helioscopes, like the proposed  AMELIE (An Axion Modulation hELIoscope Experiment), 
which could be sensitive to ALPs with masses from a few $\rm{meV}$ to several $\eV$, 
thanks to the use of a Time Projection Chamber~\cite{Galan:2015msa}. 
Studies for low mass WIMPs (Weak Interacting Massive Particles) are already being carried out by the TREX-DM experiment~\cite{Irastorza:2015dcb,Castel:2018gcp},
which is taking data at the Canfranc Underground Laboratory (LSC)~\cite{Castel:2019ngt}. 
The project aims at demonstrating the feasibility to reach low backgrounds at low energy thresholds for dark matter searches, 
which require similar detection conditions as for ALPs.

%...........................
%...................................

%However, a dedicated investigation is necessary to assess if such a detector might reach the $m_a \sim 10 \;  \eV$ range in a range of energy lower than CAST.

Concerning the ALP flux coming from resonant TP-ALP conversions in the radiative zone of the Sun, corresponding to an axion mass 
$m_a \sim 130 \; \eV$ (short-dashed curve), the flux is expected to be much larger than the Primakoff one above the CAST threshold. However, CAST cannot detect it due to the loss of coherence of ALP-photon conversions in the detector. In principle, ALPs with mass $m_a \sim 100 \; \eV$ could be detected with a dark matter detector like the Cryogenic Underground Observatory for Rare Events (CUORE), which exploits the inverse Bragg-Primakoff effect to detect solar axions~\cite{Li:2015tsa}. CUORE is expected to cover a mass range $m_a \lesssim 100 \; \eV$. 
Notice,  that for $m_a \geq 10 \; \eV$ there are some cosmological constraints to be taken into account. 
Indeed, the ionization of primordial hydrogen ($x_{\rm{ion}}$) from ALPs decaying into photons  sets the bound $g_{a \gamma} \lesssim 5 \times 10^{-13} \; \GeV^{-1} \; \textrm{for} \; m_a \sim 10^2 \; \eV$~\cite{Cadamuro:2011fd}.
{ However, in cosmological models with low-reheating temperature these bounds can be easily evaded 
(see, e.g.~\cite{Jaeckel:2017tud}, for a discussion). On the contrary, our ALP signal from the Sun is not affected by the cosmological model.
Therefore, its possible detection would also point towards a nonstandard cosmological scenario.}

Moving now to the case of the non-resonant conversions of TP modes ($m_a=0$, long-dashed curve) we realize that this contribution can reach { a few $\%$} of the Primakoff process in the same energy
range. In case of a positive detection of an ALP flux in a future helioscope, precision spectral studies in principle might determine it as an excess with respect to the expected
Primakoff flux. 

Finally, the case of the  ALP flux from LP-ALP conversions (dot-dashed curve) has been recently discussed in Ref.~\cite{OHare:2020wum}. 
There, the authors suggest the possibility of detecting ALPs from LP conversions in the energy range $10^{-2} \; \keV \lesssim \omega \lesssim 10^{-1} \; \keV$ through an upgraded version of IAXO~\cite{Armengaud:2019uso}. They forecast to have a sensitivity down to $m_a \simeq 10^{-2} \; \eV$.
We address the interested reader to this interesting and detailed work {for further details}.

%%%%%%%%%%%%%%%%%%%
\section{Conclusions}
\label{sec:conclusions}
%%%%%%%%%%%%%%%%%%%

We have characterized the ALP production in the large-scale solar magnetic fields and discussed the perspectives for their detection in helioscopes and dark matter detectors. 
In particular, we have characterized both resonant and non-resonant  conversions of transverse photons, 
which have not been taken into account so far. 
At this regard, we have considered realistic models for the solar $B$-field in the radiative zone and in the tachcoline of the Sun. 
We first studied the problem from a theoretical point of view, using a kinetic approach based on the evolution of the density matrix 
for the photon-ALP ensemble. With this approach, we estimated the production rate of ALPs in the Sun 
and we used it to estimate the ALP flux expected at Earth. 
The expression of the ALP production rate obtained in this way is completely general 
and has been specialized to study ALPs production from both LP and TP conversions.

In the case of the ALP flux from  LP-ALP conversions, 
we reproduce the result recently obtained in~\cite{Caputo:2020quz}, using the thermal quantum field theory approach.
This flux results  peaked at $E\sim 100$ eV, and might be detectable with an upgraded version of IAXO~\cite{OHare:2020wum}. 
The ALP flux from TP conversions,
for ALPs with mass $m_a \sim 10 \; \eV$, associated to resonant conversions in the tachochline, is found to be peaked below the CAST threshold. 
A dedicate investigation is necessary to assess the experimental possibility to dectect such a low-energy flux. Conversely, the ALP flux arising from transverse photon-ALP conversions for ALPs with mass $m_a \sim 100 \; \eV$ in the radiative zone, is dominant above the CAST threshold and it is larger that the Primakoff one. In principle, this flux might be detected using the dark matter detector CUORE.
Furthermore, this ALP flux allows to improve the bound on $g_{a \gamma}$ from energy-loss in the Sun, even exceeding
 the bound from Helium-burning stars in Globular Clusters.
The ALP flux from non-resonant conversions of TP modes can reach { a few $\%$} of the Primakoff process in the same energy
range. Therefore, it might produce a distortion of this flux, possibly producing observable signatures in the case of a
precise measurement of the solar ALP spectrum.

In conclusion, our work completes the recent studies of Ref.~\cite{Caputo:2020quz,OHare:2020wum}  
about the production of ALPs in the solar magnetic fields via longitudinal plasmons, 
including also the analysis of the photon transverse mode. 
Despite challenges in measuring this flux, it is intriguing to realize that the Sun can be the source of  additional ALP fluxes 
beyond the well-studied one from Primakoff conversions. 
A positive measurement of this flux would shed new light not only on ALPs, but also on the magnetic field in the Sun.

\section*{Acknowlegments}
We warmly thank Andrea Caputo, Hugo Tercas, Giuseppe Lucente, Joerg Jaeckel and Lennert Thormaehlen for valuable comments on our manuscript. 
The work of P.C. and 
A.M. is partially supported by the Italian Istituto Nazionale di Fisica Nucleare (INFN) through the ``Theoretical Astroparticle Physics'' project
and by the research grant number 2017W4HA7S
``NAT-NET: Neutrino and Astroparticle Theory Network'' under the program
PRIN 2017 funded by the Italian Ministero dell'Universit\`a e della
Ricerca (MUR).

\section*{Appendix A: Photon-ALP mixing in a plasma}
\label{sec:coupling}
Let us now consider an ALP-photon system in a plasma. We start from the Lagrangian of a photon coupled with the pseudoscalar field $a$, i.e. the ALP field
%%%%
\begin{equation}
\mathcal{L}= -\frac{1}{4}F_{\mu \nu}F^{\mu \nu} + \frac{1}{2}(\partial_{\mu}a \partial^{\mu} a -m^2_a a^2)- \frac{1}{4} g_{a \gamma} a F_{\mu \nu} \tilde{F}^{\mu \nu}+ J_\mu A^\mu \; ;
\label{eq:coupledlagrangian}
\end{equation} 
%%%%
where $g_{a \gamma}$ is the axion-photon coupling, $J_\mu$ is the electromagnetic current, $A^{\mu}$ is the vector potential, $F_{\mu \nu}$ is the electromagnetic field tensor and $\tilde{F}_{\mu \nu}= 1/2 \epsilon_{\mu \nu \rho \sigma} F^{\rho \sigma}$ its dual.
From Eq.~(\ref{eq:coupledlagrangian}) one recovers Maxwell's equations

%%%%
\begin{eqnarray}
 \partial_{\mu} F^{\mu \nu}&=&J^\nu + g_{a \gamma}\tilde{F}^{\mu \nu} \partial_\mu a \,\  \nonumber \\
( \Box + m^2_a)a &=& -\frac{1}{4} g_{a \gamma} F_{\mu \nu} \tilde{F}^{\mu \nu} \,\ \nonumber \\
\partial_\mu \tilde{F}^{\mu \nu} &=& 0 \; .
\label{eq:maxw}
\end{eqnarray}
%%%%

%....................................................
\subsection*{Transverse modes}
%....................................................
Transverse modes are characterized by an electric field ${\bf E}$ transverse to the photon momentum and a magnetic field ${\bf B}$ transverse to both. 
We consider a strong external magnetic field ${\bf B}_{\rm{ext}}$ such that the total field is ${\bf B} \approx {\bf B}_{\rm{ext}}$. 
According to the discussion of Section \ref{sec:dispersion} the photon dispersion relation for TP is
%%%%
\begin{equation}
\omega^2= k^2+ \omega^2_{\rm {p}} \; ;
\end{equation}
%%%%
where $\omega_{\rm p}$ is the plasma frequency. 
 For the purpose of our discussion we rewrite just three Maxwell's equations [Eqs.~(\ref{eq:maxw})] in a non-explictly covariant form 
%%%%
\begin{eqnarray}
\nabla \cdot {{\bf E}} &=& \rho_e- g_{a \gamma} {{\bf B}_{\rm{ext}}}\cdot \nabla a \,\ , \nonumber  \\
\nabla \times {{\bf B}}_{\rm{ext}} -\partial_t {\bf E} &=& -g_{a \gamma} {{\bf B}}_{\rm{ext}} \partial_t a+ {\bf J}\,\ , \nonumber  \\
(\Box + m^2_a) a &=& - g_{a \gamma} {{\bf B}}_{\rm {ext}} \cdot \partial_t {\bf A}\;\ ,
\label{eq:max2}
\end{eqnarray}
%%%%
where $\rho=-en_e$ is the electron charge density, ${\bf A}$ is the time-varying part of the vector potential for the external magnetic field and $ \Box = \partial_t^2 - \nabla^2$. Note that we are considering only the electrons in plasma equations, since we are assuming that ions provide a uniform background which does not partecipate in plasma motion. 
For the transverse mode $ {\bf k} \cdot {{\bf B}}_{\rm{ext}}=0$. Thus for clarity of notation we identify the magnetic field ${{\bf B}}_{\rm{ext}}\equiv {\bf B}_{\rm T}$, to denote that it is a \textit{transverse} field. Moreover, we make the assumption that ${\bf E} \ll {\bf B}_{\rm T}$. 
Then, Maxwell's equations [Eqs.~(\ref{eq:max2})] become
%%%%
\begin{eqnarray}
 \Box {\bf A} &=& g_{a \gamma} {\bf B}_T \partial_t a \,\ , \nonumber \\
 (\Box+m^2_a) a &=&- g_{a \gamma} {\bf B}_T \cdot \partial_t {\bf A} \,\ .
 \end{eqnarray}
%%%%
We specialize our calculation to a wave of frequency $ \omega $ propagating in the $z$-direction and we denote with $A_\bot$ and $A_\parallel$ the components of the vector potential ${\bf A}$ perpendicular and parallel to ${\bf B}_{T}$ respectively.
Thus the equations of motion for the TP-ALP system become \cite{Raffelt:1987im}
%%%%
\begin{equation}
\biggl[\omega^2+ \partial_z^2 + 2 \omega^2 
\begin{pmatrix}
\Delta_\bot / \omega & n_R & 0 \\
n_R & \Delta_\parallel/  \omega  & g_{a \gamma} B_T / 2\omega \\
0 & g_{a \gamma} B_T / 2\omega & -m^2_a/2 \omega^2
\end{pmatrix}
\biggr]
\begin{pmatrix}
A_\bot \\
A_\parallel \\
a
\end{pmatrix}
=0 \; ;
\label{eq:mottrasv}
\end{equation}
%%%%
where $n_R$ corresponds to the so called \textit{Faraday effect}, which denotes the possibility of rotation of the plane of polarization in optically active media with a consequent mixing of $A_\bot $ and $A_\parallel $. Moreover
%%%%
\begin{eqnarray}
 \Delta_\bot &=& \Delta_{\rm{p}}+ \Delta_\bot^{\rm{CM}}\,\ , \nonumber \\
 \Delta_\parallel &=& \Delta_{\rm{p}}+ \Delta_\parallel^{\rm{CM}} \,\ , \nonumber \\
 \Delta_{\rm{p}} &=& -\frac{\omega_{\rm p}}{2 \omega}  \; .
\end{eqnarray}
%%%%
The terms $\Delta_{\bot , \parallel}^{\rm{CM}}$ describe the Cotton-Mouton effect, i.e. the birifrangence of fluids in the presence of a transverse magnetic field. The vacuum Cotton-Mouton effect arises from QED one-loop corrections to the photon polarization when an external magnetic field is present. In this case we define $\Delta_{\rm{QED}}=|\Delta_\bot^{\rm{CM}}-\Delta_\parallel^{\rm{CM}}|$ and it is defined as
%%%%
\begin{equation}
\Delta_{\rm{QED}}= \frac{24 \alpha^2}{135}\frac{\rho_B}{m_e^4}\omega\; .
\end{equation}
%%%%
This QED correction is found to be negligible with respect to $\Delta_{\rm p}$ in the case of solar plasma. 
For transverse modes only the component $A_\parallel$ of the vector potential couples to the ALP. Moreover, if we neglect the Faraday effect ($n_R=0$) and we assume that the magnetic field is uniform, we can reduce the general $3 \times 3$ problem of Eq.~(\ref{eq:mottrasv}) to the $2 \times 2$ system involving only $A_\parallel$ and $a$. In the ultrarelativistic limit, i.e. for energies $\omega \gg m_a$ and $\omega \gg \omega_{\rm p}$, we can linearize Eq.~(\ref{eq:mottrasv}). As a result of linearization we obtain a linear Schr${\rm{\ddot o}}$dinger-like equation \cite{Raffelt:1987im}
%%%%
\begin{equation}
i \partial_z 
\begin{pmatrix}
A_\parallel \\
a
\end{pmatrix}
= 
-\omega I +
\begin{pmatrix}
{\omega^2_p}/{2 \omega} & g_{a \gamma}B_T /2 \\
g_{ a \gamma} B_T /2 & {m^2_a}/{2 \omega}
\end{pmatrix}
\begin{pmatrix}
A_\parallel \\
a
\end{pmatrix} \; .
\end{equation}
%%%%
Thus the Hamiltonian for the transverse modes reads (up to an overall phase diagonal term)
%%%%
\begin{equation}
H_{T}= 
\begin{pmatrix}
{\omega^2_p}/{2 \omega} & g_{a \gamma} B_T /2 \\
g_{a \gamma} B_T /2 & {m^2_a}/{2 \omega}
\end{pmatrix} \; .
\label{eq:htrans}
\end{equation}
%%%%
Finally, we obtain the TP-ALP conversion probability after traveling a distance $z$ in a uniform magnetic field $B_T$ \cite{Raffelt:1987im}
%%%%
\begin{equation}
P(\gamma_T \rightarrow a) = (\Delta_{a \gamma}^T z)^2 \frac{\sin^2(\Delta^T_{\rm {osc}} z/2)}{(\Delta^T_{{\rm osc}} z/2)^2} \; ;
\label{eq:probconv}
\end{equation}
%%%%
where we have introduced
%%%%
\begin{eqnarray}
 \Delta_{a \gamma}^T &=& g_{a \gamma} B_{T}/2 \,\ , \nonumber \\
 \Delta^T_{\rm {osc}} &=& \sqrt{4 {\Delta_{a \gamma}^T}^2+\biggl( \frac{\omega^2_{\rm p}-m^2_a}{2 \omega} \biggr)^2} \; .
\end{eqnarray}
%%%%

%....................
\subsection*{Longitudinal modes}
%..........................
Longitudinal modes are allowed only in presence of a medium, that is in our discussion is the solar plasma. The photon dispersion relation for LP is
%%%%
\begin{equation}
\omega^2= \omega^2_{\rm p}\; .
\end{equation}
%%%%
In this case $\nabla \cdot {{\bf B}} \neq 0$, thus  the plasma equations of motion and the relevant Maxwell's equations are \cite{Tercas:2018gxv,Mendonca:2019eke,Das:2004ee}
%%%%
\begin{align}
& \frac{\partial n_e}{\partial t}+ \nabla \cdot (n_e {\bf v})=0 \; ; \label{eq:l1} \\
& \frac{\partial {\bf v}}{\partial t}+ ({\bf v} \cdot {\bf \nabla}){\bf v}= \frac{e}{m} \biggl( {\bf E} + {\bf v} \times {\bf B} \biggr) - \frac{1}{m n_e} \nabla p \; ; \label{eq:l2} \\
& \nabla \cdot ({\bf{E}}+ g_{a \gamma} {{\bf B}} a)=e(n_e-n_e^0) \; ;\label{eq:l3} \\
& (\Box + m_a^2) a= g_{a \gamma} a \bf{E} \cdot \bf{B}\; \label{eq:l4} .
\end{align}
%%%%
The goal is now compute the LP-ALP conversion probability, assuming that no LP absorption exists in the plasma. In the case of longitudinal modes we have to combine Eqs.~(\ref{eq:l1}) and (\ref{eq:l2}) with Maxwell's equations. We consider a uniform external magnetic field along the $z$-direction ${\bf B}= B_L {\hat z}$ and we consider the plane wave approximation, i.e. we assume that all the fields vary as $e^{i({\bf k} \cdot {\bf x} - \omega t)}$. Moreover, we take into account a small perturbation to the electron number density
%%%%
\begin{equation}
n_e=n_e^0+ \delta n \; ;
\end{equation}
%%%%
where $n_e^0$ is the equilibrium value and $\delta n \ll n_0$. If we finally take $\omega= \omega_{\rm {p}}= \omega_{a}$, i.e. the photon energy coincident with the plasma frequency and with the ALP energy, we can linearize Maxwell's equations Eqs.~(\ref{eq:l3})-(\ref{eq:l4}) for longitudinal modes,
obtaining  \cite{Tercas:2018gxv}
%%%%
\begin{equation}
i \partial_z
\begin{pmatrix}
A_L\\
a_k
\end{pmatrix}
=
\begin{pmatrix}
\omega_p & {g_{a \gamma B_L}}/{2} \\
{g_{a \gamma B_L}}/{2} & \omega_a
\end{pmatrix}
\begin{pmatrix}
A_L\\
a_k
\end{pmatrix} \; ;
\end{equation}
%%%%
where we have introduced the fields $\delta n=i n_e A_L$ and $a=\omega_{\rm{p}} m_e a_k/ek$.
Thus the Hamiltonian of the system is 
%%%%
\begin{equation}
H_L=
\begin{pmatrix}
\omega_p & \Delta_{a \gamma}^L \\
\Delta_{a \gamma}^L & \omega_a
\end{pmatrix} \; .
\label{eq:hlong}
\end{equation}
%%%%
The LP-ALP conversion probability is 
%%%%
\begin{equation}
P(\gamma_{\rm{LP}} \rightarrow a) = (\Delta_{a \gamma}^L z)^2 \frac{\sin^2(\Delta^L_{{\rm osc}} z/2)}{(\Delta^L_{\rm {osc}} z/2)^2} \; ;
\label{eq:lp}
\end{equation}
%%%%
where 
%%%%
\begin{align}
& \Delta_{a \gamma}^L= \frac{g_{a \gamma} B_L}{2}\; ; \\
& \Delta^L_{\rm{osc}}=\sqrt{4 {\Delta_{a \gamma}^L}^2+ ( \omega_a- \omega_{\rm p})^2} \; .
\end{align}
%%%%

%%%%%%%%%%%%%%%%%%%%%%
\section*{Appendix B: Kinetic Approach}
%%%%%%%%%%%%%%%%%%%%%%

In order to determine the ALP emission rate in the Sun due to the conversions of photons in magnetic fields we closely follow the kinetic approach
developed in \cite{Redondo:2013lna} for the case of hidden photons.
We consider the system introduced in Sec. \ref{sec:rate}
constituted by two bosonic fields $A$ and $S$. We assume that the field $A$ interacts with the medium, namely with the solar plasma, which can absorb a quantum with rate $\Gamma_{\rm{abs}}$ and produce one with rate $\Gamma_{\rm{prod}}$. 
The kinetic equation for the density matrix $\rho$ of the $A-S$ ensemble is given by
 \cite{Sigl:1992fn}
%%%%
\begin{equation}
\dot{\rho}=-i[\Omega, \rho]+\frac{1}{2} \{G_{\rm{prod}}, I+ \rho \}-\frac{1}{2} \{G_{\rm {abs}}, \rho \} \; ;
\label{eq:liouville}
\end{equation}
where
%%%%
\begin{equation}
\Omega=
\begin{pmatrix}
\omega_A & \mu \\
\mu & \omega_S 
\end{pmatrix}
= \frac{\omega_A+\omega_S}{2} I +
\begin{pmatrix}
\frac{1}{2}\Delta \omega & \mu \\
\mu & -\frac{1}{2} \Delta \omega
\end{pmatrix} \; .
\end{equation}
%%%%
and
%%%%
\begin{equation}
\begin{split}
G_{\rm {prod}}=
\begin{pmatrix}
\Gamma_{\rm{prod}} & 0 \\
0 & 0
\end{pmatrix} \; ; \\
G_{\rm{abs}}=
\begin{pmatrix}
\Gamma_{\rm{abs}} & 0 \\
0 & 0
\end{pmatrix} \; \; . 
\end{split}
\end{equation}
%%%%
In thermal equilibrium $\Gamma_{\rm{prod}}= e^{-\omega/T} \Gamma_{\rm{abs}}$ and the $S$  type particles are not excited, while we assume that the $A$ type particles obey the Bose-Einstein statistics $f_{\rm{BE}}=(e^{\omega/T}-1)^{-1}$.
 A non equilibrium situation of Eq. (\ref{eq:liouville})  is described with a small deviation $\delta \rho$ from the thermal equilibrium state $\rho_{\rm{eq}}$, thus 
%%%%
\begin{equation}
\rho=\rho_{\rm{eq}}+ \delta \rho= 
\begin{pmatrix}
f_{\rm{BE}} & 0 \\
0 & 0 
\end{pmatrix}
+ \delta \rho \; .
\label{eq:ro}
\end{equation}
%%%%
In Eq.~(\ref{eq:liouville}) the collision terms, i.e. the anticommutators, vanish for $\rho_{\rm{eq}}$, thus the Liouville equation reduces to
%%%%
\begin{equation}
\dot{\rho}=-i[\Omega, \rho]- \frac{1}{2} \{G, \delta \rho \} \; ;
\label{eq:liouvillenew}
\end{equation}
%%%%
where we have introduced $G=\diag(\Gamma, 0)$, with $\Gamma=(1-e^{-\omega/T})\Gamma_{\rm{abs}}$, i.e. the total collisional rate. We can write $\delta \rho$ as
%%%%
\begin{equation}
\delta \rho=
\begin{pmatrix}
n_A & g \\
g & n_S
\end{pmatrix}\; ;
\label{eq:deltaro}
\end{equation}
%%%%
where $n_A$ is the occupation numbers of $A$ quanta, $n_S$ the occupation numbers of $S$ quanta and $g$ represents the mixing between the two levels.
If we insert Eqs.~(\ref{eq:ro})--(\ref{eq:deltaro}) into Eq.~(\ref{eq:liouvillenew}) we obtain the equations of motion
%%%%
\begin{align}
&\dot{n}_A= -\Gamma n_A-2\mu \textrm{Im}(g) \; ; \label{eq:na}  \\
& \dot{n}_S= 2 \mu \textrm{Im}(g) \; ; \label{eq:ns} \\
& \dot{g}= -\bigl( \frac{1}{2} \Gamma + i \Delta \omega \bigr)g +i \mu ( f_{\rm{BE}}+ n_A-n_S) \; . \label{eq:g}
\end{align}
%%%%
The mixing $\mu$ is always small so basically we are never far from the thermal equilibrium, i.e. $f_{\rm{BE}} \gg n_A$ and $f_{\rm{BE}} \gg n_S$. In this limit Eq.~(\ref{eq:g}) reads
%%%%
\begin{equation}
\dot{g}=-\bigl( \frac{1}{2} \Gamma + i \Delta \omega \bigr) g+ i \mu f_{\rm{BE}} \; .
\label{eq:gnew}
\end{equation}
%%%%
Assuming the initial condition $g(0)=0$ the solution of Eq.~(\ref{eq:gnew}) is then
%%%%
\begin{equation}
g(t)= \frac{1- e^{-(i \Delta \omega+ \Gamma/2)t}}{\Delta \omega-i \frac{\Gamma}{2}} \mu f_{\rm{BE}}\; .
\label{eq:sol}
\end{equation}
%%%%
After an initial transient, Eq.~(\ref{eq:sol}) approaches the steady-state solution 
%%%%
\begin{equation}
g_{\infty}=\frac{\Delta \omega+i \Gamma/2}{\Delta \omega^2 + \Gamma^2/4 } \mu f_{\rm{BE}}\; .
\label{eq:solstaz}
\end{equation}
%%%%
If we insert this latter in Eq.~(\ref{eq:ns}) we finally obtain the $S$ quanta production rate 
%%%%
\begin{equation}
\Gamma_S^{\rm{prod}} \equiv \dot{n}_S=  \frac{\Gamma \mu^2}{(\omega_A-\omega_S)^2 + \Gamma^2/4} \frac{1}{e^{\omega/T}-1} \; .
\label{eq:rategenerale}
\end{equation}
%%%%
From Eq.~(\ref{eq:rategenerale}) we obtain that the $A-S$ mixing process is \textit{resonant}, i.e. it is maximal for $\omega_A=\omega_S$. 
The result in Eq.~(\ref{eq:rategenerale}) is completely general and it is valid both at the resonance and off-resonance, since it has been obtained on the only assumption that the mixing term $\mu$ is small relative to the diagonal terms $\sim \Delta \omega$. This condition always applies in the solar plasma, both for photon TP  and LP modes.
 
%%%%%%%%%%%%%%%%%%%%%%
\section*{Appendix C: Thermal Field Theory Approach}
%%%%%%%%%%%%%%%%%%%%%%

Here, we show how the equations for the axion production rate  
can also be derived using the thermal field theory approach proposed in Ref.~\cite{Weldon:1983jn}.
In this formalism, the production rate can be expressed as
\begin{align}
\label{eq:}
\Gamma_{\rm prod}=-\frac{{\rm Im~}\Pi_a}{\omega\left( e^{\omega/T}-1 \right)} \,,
\end{align}
where the axion self-energy $\Pi_a$ is defined in terms of the exact axion propagator $D_a$ through 
$ D_a^{-1} =(k^2+ \Pi_a^2)$.
Making use of the axion-photon interaction, 
$L_{a\gamma} = 
 g_{a\gamma}  (\partial_\mu a) A_\nu \widetilde{F}_{\rm ext}^{\mu\nu}$,
it is possible to 
formally express $\Pi_a$ in terms of the exact photon propagator $D^{(\gamma)}$:
\begin{align}
\label{Eq:axion_polarization_tensor}
\Pi_a = m_a^2 +  g_{a\gamma}^2 
p_\mu \,p_\nu \widetilde{F}_{\rm ext}^{\mu\alpha} \, \widetilde{F}_{\rm ext}^{\nu\beta} \,
D^{(\gamma)}_{\alpha\beta} \,,
\end{align}
where $p$ is the axion momentum. 
Thus, the problem reduces to the determination of the photon propagator in a plasma at finite temperature and in presence of an external magnetic field. 
This problem is, in general, quite difficult (see, e.g., \cite{Pal:2020bwo}).
However, in the present case we see that the magnetic field corrections to the propagator are small. 
In fact, the magnetic field insertions are of the order of 
$\sqrt{eB_{\rm ext}} = 7.7\,B_1^{1/2} \, {\rm eV}$, where $B_1=B/1$T.
This scale is always much lower than the temperature in the plasma, for the conditions we are interested in. 
We thus ignore these corrections and consider the photon propagator expected in an isotropic medium at finite temperature.
This case is considerably simpler and is discussed in several references
(see, e.g., Ref.~\cite{Raffelt:1996wa} for a pedagogical introduction).
%Here, we strictly follow Ref.~\cite{Weldon:1982aq}.

The photon polarization tensor in and isotropic plasma can be expressed in a covariant form 
introducing the plasma four-velocity $ u^\mu $, which in the plasma frame reduces to $ (1,\vec 0) $, 
and the photon momentum vector $ K^\mu$ which in the plasma frame reduces to  $=(\omega, \vec k) $.
Following Ref.~\cite{Weldon:1982aq}, we define the tensor
\begin{align}
\widetilde \eta_{\mu\nu}=\eta_{\mu\nu}-u_\mu u_\nu  \,, 
\end{align}
and the two vectors
\begin{align}
&\widetilde K_{\mu}=K_{\mu}-\omega u_{\mu}  \,, \\
& q_\alpha = \frac{1}{\sqrt{K^2 k^2}}
\left( k^2 u_\alpha +\omega \widetilde K_\alpha \right) \,.
\end{align}
Notice that $q_\alpha \to (\sqrt{\omega^2-k^2})^{-1} (|\vec k|,\omega \hat k)$
%%
%\begin{align}
%\label{eq:epsilon_L}
%q_\alpha \to \frac{1}{\sqrt{\omega^2-k^2} }(|\vec k|,\omega \hat k)\,
%\end{align}
%%
in the plasma reference frame, and so it represents the unit vector along the longitudinal direction
(see, for example,~\cite{An:2013yfc}).\footnote{The temporal component is chosen in such a way to make it perpendicular to $K$, which, because of gauge invariance, is necessarily an eigenvector of the photon self-energy.}
Hence, we can define the projection operator along the longitudinal direction as
\begin{align}
Q_{\alpha \beta}=-\frac{1}{K^2 k^2}\left( k^2 u_\alpha +\omega \widetilde K_\alpha \right) 
\left( k^2 u_\beta +\omega \widetilde K_\beta \right) \,.
\end{align}
The orthogonal projector is defined as 
\begin{align}
\label{eq:}
 P_{\alpha \beta}=\widetilde{\eta}_{\alpha \beta}+\frac{\widetilde K_{\alpha} \widetilde K_{\beta}}{k^2} \,.
\end{align}
Notice that $P$ and $Q$ satisfy the properties $ P^2=P $, $ Q^2=Q $, and $ (PQ)^\mu_\alpha =(QP)^\mu_\alpha =0$.

With these definitions, the exact photon propagator in an isotropic medium has the form 
\begin{align}
D^{(\gamma)}_{\alpha \beta}= -\frac{P_{\alpha\beta}}{K^2-\pi_T}
-\frac{Q_{\alpha\beta}}{K^2-\pi_L}
+(\alpha-1)\frac{K_{\alpha}K_{\beta}}{K^4}\,,
\end{align}
where $\pi_{T,L}$ are eigenvalues of the photon self-energy corresponding to the transverse and longitudinal direction, while 
$ \alpha $ is the gauge parameter.

We can now proceed to extract the tensorial form of the axion self-energy from Eq.~\eqref{Eq:axion_polarization_tensor}.
First, let's notice that $K=p+q_t$, where $p$ is the axion momentum and $q_t=(0,\vec q_t)$ is the momentum transferred from the magnetic field. 
The energies of the axion and of the photon are the same. 
Moreover, the axion and photon momenta have the same direction. 
Thus, we can calculate the momentum transfer as $|\vec q_t|=|\vec q_\gamma|-|\vec q_a|
=\sqrt{\omega^2-m_\gamma^2}-\sqrt{\omega^2-m_a^2}$.
Since $|\vec q_a|\sim \omega$, we can assume that $\vec p\sim \vec k$, in our conditions. 
With this simplification, the term
$p_\mu \widetilde{F}_{\rm ext}^{\mu\alpha}$, which appears in Eq.~\eqref{Eq:axion_polarization_tensor}, reduces to
$(-\vec k\cdot \vec B, -\omega \vec B)$ in the plasma frame.
%\footnote{Notice, that in the plasma frame, 
%\begin{align}
%\widetilde{F}_{\rm ext}^{\mu\nu}  = 
%\begin{pmatrix}
%0 & -B_x & -B_y & -B_z \\
%B_x & 0 & 0 & 0 \\
%B_y & 0 & 0 & 0\\
%B_z & 0 & 0 & 0
%\end{pmatrix}\,.
%\end{align}
%}
Thus, we find $p_\mu \widetilde{F}_{\rm ext}^{\mu\alpha}q_\alpha =\sqrt{K^2}B_{T}$, and
$ p_\mu \widetilde{F}_{\rm ext}^{\mu\alpha} p_\mu \widetilde{F}_{\rm ext}^{\mu\beta} P_{\alpha\beta} 
= \omega^2 \left( B^2-B_T^2 \right) = \omega^2 B_ L^2$,
where $B_{L,T}$ refer, respectively, to the direction parallel and perpendicular to the photon momentum. 
Thus, 
\begin{align}
\Pi_{a,T}= g_{a\gamma}^2 \frac{\omega^2\,B_T^2}{K^2-\pi_T} \,, ~~
\Pi_{a,L}= g_{a\gamma}^2 \frac{K^2\,B_L^2}{K^2-\pi_L}\,.
\end{align}

We can now proceed to extract the imaginary parts.
For the transverse mode we find:
\begin{align}
\label{eq:}
{\rm Im}~\Pi_{a,T}=g_{a\gamma}^2 \omega^2\,B_T^2\frac{1}{K^2-{\rm Re ~}\pi_T-i \,{\rm Im~}\pi_T} \,.
\end{align}
In a non-relativistic plasma, ${\rm Re}~\pi_T=  \omega_p^2$.
Moreover, we should interpret ${\rm Im~}\pi_T=-\omega \Gamma_T$.
So, finally,
\begin{align}
\label{eq:}
{\rm Im}~\Pi_{a,T}=-g_{a\gamma}^2 \omega^2\,B_T^2
\frac{ \omega \Gamma_T}{\left( m_a^2-\omega_p^2 \right)^2 +\left(  \omega \Gamma_T \right)^2}\,,
\end{align}
and so,
\begin{align}
\label{eq:}
\Gamma_{a,T}=\frac{g_{a\gamma}^2 \omega^2\,B_T^2 \, \Gamma_T}
{\left[ \left( m_a^2-\omega_p^2 \right)^2 +\left(  \omega \Gamma_T \right)^2 \right]\left( e^{\omega/T}-1 \right)}\,.
\end{align}

For the longitudinal mode we find:
\begin{align}
\label{eq:}
{\rm Im}~\Pi_{a,L}=g_{a\gamma}^2 K^2\,B_L^2\frac{1}{K^2-{\rm Re ~}\pi_L-i \,{\rm Im~}\pi_L} \,.
\end{align}
In a non-relativistic plasma, 
$ {\rm Re}~\pi_{L}= K^2 \omega_p^2/\omega^2 $.
Thus, multiplying numerator and denominator by $(\omega^2/K^2)$, we find
\begin{align}
\label{eq:}
{\rm Im}~\Pi_{a,L}=g_{a\gamma}^2 K^2\,B_L^2\frac{\omega^2/K^2}{\omega^2-\omega_p^2
-i (\omega^2/K^2){\rm Im~}\pi_L}\,.
\end{align}
We should therefore interpret $(\omega^2/K^2){\rm Im~}\pi_L=-\omega \Gamma_L$.
So, finally,
\begin{align}
\label{eq:}
{\rm Im}~\Pi_{a,L}=-g_{a\gamma}^2 \omega^2\,B_L^2
\frac{ \omega \Gamma_L}{\left( \omega^2-\omega_p^2 \right)^2 +\left(  \omega \Gamma_L \right)^2}\,,
\end{align}
and 
\begin{align}
\label{eq:gammalth}
\Gamma_{a,L}=\frac{g_{a\gamma}^2 \omega^2\,B_L^2 \, \Gamma_L}
{\left[ \left( \omega^2-\omega_p^2 \right)^2 +\left(  \omega \Gamma_L \right)^2 \right]\left( e^{\omega/T}-1 \right)}\,.
\end{align}

\end{document}